\documentclass[11pt]{article}

\usepackage{graphicx}
\usepackage[a4paper, margin=2.3cm]{geometry}
\usepackage[backend=biber,style=ieee,natbib=true]{biblatex}
\usepackage[dvipsnames]{xcolor}
\usepackage{booktabs}
\usepackage{tabularx}
\usepackage{ragged2e}
\newcolumntype{Y}[1]{>{\hsize=#1\hsize\RaggedRight\arraybackslash}X}\usepackage{acro}
\usepackage{amsmath}
\usepackage{amssymb}
\usepackage[export]{adjustbox}
\usepackage{subcaption}
\usepackage{tikz}
\usetikzlibrary{positioning,arrows.meta,fit,backgrounds,calc}
\usepackage{float}
\usepackage{siunitx}
\usepackage[hidelinks]{hyperref}
\usepackage{authblk}

\usepackage[T1]{fontenc}

\DeclareAcronym{SUMO}{short=SUMO, short-plural=, long=Simulation of Urban MObility, long-plural=}
\DeclareAcronym{OSM}{short=OSM, short-plural=, long=OpenStreetMap, long-plural=}
\DeclareAcronym{AQI}{short=AQI, short-plural=AQIs, long=Air Quality Index, long-plural=Air Quality Indices}
\DeclareAcronym{PDE}{short=PDE, long=partial differential equation, list=Partial Differential Equation}
\DeclareAcronym{FDM}{short=FDM, long=finite difference method, list=Finite Difference Method}
\DeclareAcronym{CFL}{short=CFL, short-plural=, long=Courant--Friedrichs--Lewy, long-plural=}
\DeclareAcronym{HBEFA}{short=HBEFA, short-plural=, long=Handbook Emission Factors for Road Transport, long-plural=}
\DeclareAcronym{PHEM}{short=PHEM, short-plural=, long=Passenger Car and Heavy-Duty Emission Model, long-plural=}
\DeclareAcronym{CFD}{short=CFD, short-plural=, long=computational fluid dynamics, long-plural=}
\DeclareAcronym{OSPM}{short=OSPM, short-plural=, long=Operational Street Pollution Model, long-plural=}
\DeclareAcronym{TraCI}{short=TraCI, short-plural=, long=Traffic Control Interface, long-plural=}
\DeclareAcronym{HC}{short=HC, short-plural=, long=hydrocarbon, long-plural=, list=Hydrocarbon}
\DeclareAcronym{PM}{short=PM, short-plural=, long=particulate matter, long-plural=}
\DeclareAcronym{NO}{short=NO, short-plural=, long=nitric oxide, long-plural=}
\DeclareAcronym{LUR}{short=LUR, short-plural=, long=land-use regression, long-plural=}
\DeclareAcronym{PEMS}{short=PEMS, short-plural=, long=portable emission measurement system, long-plural=portable emission measurement systems}
\DeclareAcronym{NABEL}{short=NABEL, short-plural=, long=Swiss National Air Pollution Monitoring Network, long-plural=Swiss National Air Pollution Monitoring Network}
\DeclareAcronym{EPA}{short=EPA, short-plural=, long=U.S. Environmental Protection Agency, long-plural=}
\DeclareAcronym{WHO}{short=WHO, short-plural=, long=World Health Organization, long-plural=}

\title{\textbf{Who Breathes the Detour? %\\[2pt]
A Coupled Traffic--Emission--Dispersion--Exposure Pipeline 
}}
 
\author{Cédric-Yséry Zeiter$^{1,2}$, Kevin Riehl$^2$, Anastasios Kouvelas$^2$}
 
\affil{
    $^1$ ETH Zürich, Department of Mathematics, Zurich, Switzerland \\
    $^2$ ETH Zürich, Institute for Transport Planning and Systems, Traffic Engineering Group, Zurich, Switzerland \\
}

\date{}
\begin{document}
\maketitle

%%%%%%%%%%%%% ABSTRACT
\begin{abstract}
Road traffic is among the largest sources of health-damaging air pollution in inhabited areas. Direct measurement of resident exposure at the scale of an entire corridor remains costly, and existing simulation approaches often target either high-fidelity local snapshots or regional averages, leaving the corridor scale underserved. We present a coupled, end-to-end pipeline that links four stages: (i) microscopic traffic simulation with SUMO, (ii) per-vehicle emissions from the HBEFA4 model, (iii) deterministic three-dimensional Eulerian diffusion solved with an implicit finite-difference scheme on a geometric vertical grid, and (iv) resident-level exposure expressed as an air quality index over a synthetic population. The design deliberately trades the fidelity of full CFD dispersion for a workflow that runs end-to-end for the residential areas of several settlements on a single laptop, and the complete implementation is released open source. We showcase the pipeline at the example of the Swiss Gotthard transit corridor, covering the residential areas of six villages, two days, four scenarios, and three pollutants, to enable assessment of congestion mitigation measures. \\ Code \& Material: \url{https://github.com/cedriczeiter/zeiter_coupled_pipeline}.
\end{abstract}

\section{Introduction}
\label{sec:intro}

% PAR 1: Roads are lifelines of settlements, but traffic harms those living nearest

Roads are the lifelines of modern settlements. Businesses and homes have historically grown along these routes, and the mobility they provide underpins much of economic life. Yet the same roads that enabled this growth now impose a severe burden on the people who live closest to them. Road traffic is among the largest sources of air pollution in inhabited areas, and the exposure it produces carries a heavy cost in disease, premature death, and diminished quality of life. Air pollution is a devastating environmental hazard that accounted for 11.62\% of all global deaths in 2019~\citep{Hu2023}. Estimates of the absolute mortality burden differ by methodology and pollutant scope: \textcite{Fuller2022} attribute 6.7 million deaths in 2019 to air pollution, a figure of 8.34 million for 2019 is reported by \textcite{Lelieveld2023}, and estimates focusing specifically on outdoor air pollution put the toll at 3.3 million premature deaths in 2010~\citep{Lelieveld2015}. In motorised economies such as the USA, Germany, and the UK, land traffic accounts for roughly one-fifth of premature mortality from outdoor air pollution~\citep{Lelieveld2015}. The very pattern of urban growth that roads made possible has placed many people in the zone of highest exposure: near busy roads, detectable increases in traffic-related pollutants reach out to about 500\,m~\citep{Boogaard2022}.

% PAR 2: The tension in the canton of Uri as a motivating example

This tension manifests sharply in the Swiss canton of Uri, through which the Gotthard corridor, one of Europe's principal north--south transit axes, passes. Its backbone is the A2, the Swiss national motorway from Basel to Chiasso. The Gotthard road tunnel on the A2 carried almost 7.1 million vehicles in 2024, more than double the volume of 1981, its first full year of operation~\citep{ASTRA2025Faktenblatt}. The A2 as a whole accumulates a disproportionate 21.7\% of all national jam-hours despite carrying only 15.9\% of motorway traffic~\citep{ASTRA2025Jahresbericht}, and the Gotthard tunnel is one of its recurrent seasonal bottlenecks. When queues form, navigation systems reroute transit traffic onto the cantonal road that passes directly through the valley's villages, blocking local buses, supply deliveries, and emergency vehicles~\citep{Campax2022}. Residents have made their frustration felt: a petition demanding protective measures collected over 3,000 signatures and was delivered to the cantonal government in February 2023~\citep{Campax2022, UrnerZeitung2023}. The canton has since deployed a package of traffic management countermeasures targeting diversion onto cantonal roads, including ramp closures at key highway entrances and speed reductions on the A2~\citep{KantonUri2026}. However, these measures address the symptom of diversion rather than the underlying exposure of residents to elevated traffic and its associated air pollution.

% PAR 3: Why measuring exposure matters

Measuring traffic-related pollution is the prerequisite for any informed response: reliable estimates of traffic emissions provide the evidence needed to design effective mitigation policy~\citep{Jang2025}. Knowing not just how much is emitted but where people are actually exposed enables planners, traffic engineers, and health authorities to target interventions where they matter most. Integrated modelling chains linking traffic activity to population exposure have been proposed to enable such policy-sensitive assessments, with the ultimate aim of mitigating the negative effects of traffic on residents' lives~\citep{HULSMANN2014732}. A crucial distinction underlies this: the difference between emissions (what leaves the tailpipe) and exposure (what people actually breathe). These two are separated by the physical transport of pollutants through air, which redistributes and dilutes emissions, causing roadside concentrations to typically fall back towards background levels within a few hundred metres of the source~\citep{Clements2009}. It is therefore exposure, and not emission alone, that determines the adverse health effects of traffic-related air pollution. Living close to heavy traffic has been linked to elevated rates of respiratory symptoms in children~\citep{VanVliet1997} and to a higher incidence of dementia in adults~\citep{Chen2017, Yuchi2020}, with the burden concentrated among those living nearest the busiest roads.

% PAR 4: Measuring exposure directly is hard

Direct measurement of exposure across an entire corridor is, however, difficult. Pollutant concentrations vary strongly over short distances~\citep{Zhu2002}. Simulations indicate that levels can differ between the two sides of a single street canyon~\citep{HULSMANN2014732}, and even quiet roads can be loaded by pollutants carried over from busier roads nearby~\citep{WOODWARD2019}. A fixed network of monitoring stations samples only a limited number of locations. Because the true spatial distribution of concentrations across the wider area remains unknown, measurements at any single station may not be representative of the surrounding environment~\citep{Duyzer2015}. The concentration at any single point depends on meteorology, terrain, and the surrounding urban geometry, all of which shape how pollutants accumulate and disperse~\citep{SILVEIRA2023119481}, making it unlikely that a sparse network of official stations can adequately capture the full range of local exposure conditions along an extended corridor.

% PAR 5: Simulation as the alternative, and the four-stage structure

Dense networks of fixed monitoring stations are expensive to operate~\citep{samake2022}, and simulation offers a complementary route that also permits counterfactual analysis, testing what would happen under different traffic management strategies before any intervention is implemented~\citep{hirschmann2010}. Coupled traffic-to-concentration assessments have reproduced the level and diurnal pattern of measured concentrations at a street-canyon station~\citep{HULSMANN2014732}, although others have been checked only against a single mobile sensor over a few hours~\citep{samake2022}. A simulation does not reproduce reality point for point. In complex terrain, grid-averaged model output and instantaneous single-point measurements cannot be expected to agree exactly~\citep{castelli}. A simulation is therefore most valuable when it is built deliberately and used with its limits in view. In practice, an integrated assessment of this kind typically comprises four linked stages. Traffic modelling establishes how many vehicles travel where, and with which driving behaviour, at macroscopic, mesoscopic, or microscopic resolution~\citep{madziel2023}. Emission modelling then converts vehicle activity into pollutant mass, distinguishing between aggregate approaches that rely on average speed and fleet composition, and instantaneous approaches that use second-by-second vehicle dynamics~\citep{Ma2012, DIB2025104791}. Dispersion modelling describes how that mass is transported through the air, yielding concentration fields~\citep{HATZOPOULOU2010315, castelli}. Finally, exposure modelling links those concentrations to the people who breathe them~\citep{HATZOPOULOU2010315}.

% PAR 6: Limitations of existing approaches

Existing simulation-based approaches, however, rarely span all four stages simultaneously: many stop at the emission stage and do not model dispersion at all, which limits their ability to assess actual exposure~\citep{Abou-Senna2013, hirschmann2010, So21042018}. Where emission inputs feed into such pipelines, average-speed traffic input remains a common baseline that smooths away the congestion and stop-and-go dynamics responsible for the largest emission peaks~\citep{Abou-Senna2013, QUAASSDORFF2016}. Beyond these methodological gaps, three further characteristics narrow the applicability of existing work. First, the coupled pipelines that do exist are themselves urban case studies, applied to cities such as Toronto~\citep{HATZOPOULOU2010315} and Riverside~\citep{zhao2024}. Second, many studies are conducted over short simulation windows rather than full days: \textcite{Abou-Senna2013} simulate a single peak hour, \textcite{zhao2024} simulate one hour, and \textcite{QUAASSDORFF2016} treat each scenario as an isolated hourly snapshot. Third, of the few pipelines that span this many stages, \textcite{zhao2024} and \textcite{HATZOPOULOU2010315} come closest to the integrated approach pursued in this work, yet neither releases a publicly available implementation, making existing approaches difficult to reproduce or transfer to a new setting. Moreover, microscale dispersion models are too computationally demanding for large-scale scenarios~\citep{HULSMANN2014732}, and microscopic traffic simulation of an entire region is computationally intensive~\citep{Tu2018}, further limiting the transferability of high-fidelity pipelines.

% PAR 7: Our method

Addressing these shortcomings requires a method that spans all four stages at once: it must retain the vehicle-level traffic detail that drives emission peaks, disperse emissions deterministically rather than stopping at the tailpipe, and cover a full simulated day. At the same time, it must stay computationally light enough to run end-to-end for the residential area of a settlement. We address these requirements with a coupled, end-to-end pipeline. Traffic is simulated microscopically in \ac{SUMO}~\citep{Lopez2018}, resolving per-vehicle positions and driving states. Emissions are then read per vehicle from \ac{SUMO}'s built-in \ac{HBEFA}4 model~\citep{sumo_hbefa4} and deposited as moving sources into a deterministic diffusion equation solved on a three-dimensional grid, in the spirit of recent grid-based \ac{PDE} approaches~\citep{samake2022, zhao2024}. The resulting concentrations are aggregated into an \ac{AQI} for every resident using a synthetic population. This reflects a deliberate trade-off: we give up some of the fidelity of full \ac{CFD} dispersion in exchange for computational tractability~\citep{zhao2024}. In return, the pipeline runs end-to-end for the residential areas of several settlements on a single laptop and remains usable in the kind of underrepresented settings that motivated the study. The entire pipeline is released as open-source software at \url{https://github.com/cedriczeiter/zeiter_coupled_pipeline}.

% PAR 8: The case study as a demonstration

We demonstrate the method on the aforementioned Gotthard corridor, where recurrent congestion at the tunnel entrance diverts traffic onto cantonal roads that pass directly through the villages of a narrow alpine valley~\citep{Campax2022}. This setting makes a compelling test case: it feeds directly into a live political discussion and, to our knowledge, has not been modelled in the scientific literature. Even though background concentrations of nitrogen dioxide (NO$_2$) and of fine particulate matter with an aerodynamic diameter of up to 2.5\,$\mu$m (PM$_{2.5}$) in Switzerland's rural and alpine areas generally remain within national limits~\citep{BAFU2025Luftqualitaet}, compliance at the regional scale does not preclude severe local hotspots~\citep{QUAASSDORFF2016}. The case study serves to show what the pipeline can resolve in such a setting.

% PAR 9: Contributions

This paper builds further on the early evidence of \textcite{zhao2024}, who demonstrated a coupled traffic-to-exposure pipeline for pedestrians at small scale, and of \textcite{HATZOPOULOU2010315}, who carried out a large-scale simulation but did not develop their own dispersion model. It makes three primary contributions:
(i)~We develop a four-stage pipeline, ranging from microscopic traffic simulation to resident exposure, and release the tool as open-source software for reproducibility and reusability.
(ii)~We decompose the \ac{AQI} score into a fixed ambient background and a traffic-attributable excess, which makes settlements and days comparable in low-background settings where the raw index is dominated by the ambient floor.
(iii)~We demonstrate the pipeline on the Swiss Gotthard corridor across six villages under low- and high-congestion conditions, illustrating how traffic volume, randomised local traffic, and dynamic rerouting show up in the resulting exposure. 

% PAR 10: Structure
The remainder of the paper reviews the relevant literature (Section~\ref{sec:litreview}), describes the method (Section~\ref{sec:methods}), introduces the Gotthard case study with its site-specific data and settings (Section~\ref{sec:casestudy}), presents results (Section~\ref{sec:results}), discusses limitations (Section~\ref{sec:discussion}), and concludes (Section~\ref{sec:conclusion}). Implementation details are collected in a technical report in Appendix~\ref{app:technical}.

\section{Literature Review}
\label{sec:litreview}

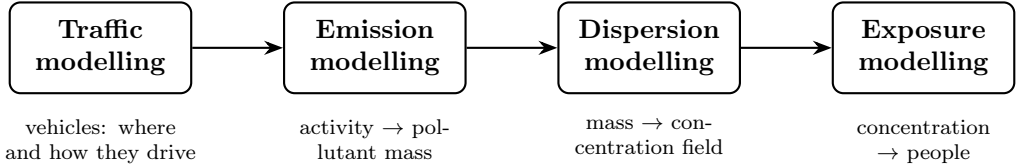
\begin{figure}[H]
\centering
\begin{tikzpicture}[
  node distance=8mm and 12mm,
  stage/.style={rectangle, rounded corners, draw, thick, align=center,
                minimum height=12mm, minimum width=24mm, font=\small\bfseries},
  lbl/.style={align=center, font=\scriptsize, text width=26mm},
  arr/.style={-{Stealth[length=2.5mm]}, thick}
]
\node[stage] (traffic) {Traffic\\modelling};
\node[stage, right=of traffic] (emission) {Emission\\modelling};
\node[stage, right=of emission] (dispersion) {Dispersion\\modelling};
\node[stage, right=of dispersion] (exposure) {Exposure\\modelling};

\draw[arr] (traffic) -- (emission);
\draw[arr] (emission) -- (dispersion);
\draw[arr] (dispersion) -- (exposure);

\node[lbl, below=2mm of traffic] {vehicles: where and how they drive};
\node[lbl, below=2mm of emission] {activity $\rightarrow$ pollutant mass};
\node[lbl, below=2mm of dispersion] {mass $\rightarrow$ concentration field};
\node[lbl, below=2mm of exposure] {concentration $\rightarrow$ people};
\end{tikzpicture}
\caption{\textbf{Four stages of the simulation pipeline}.}
\label{fig:fourstage}
\end{figure}

The literature is organised around the four stages of a simulation-based exposure assessment, shown in Figure~\ref{fig:fourstage}: traffic modelling (Section~\ref{sec:lit-traffic}), emission modelling (Section~\ref{sec:lit-emission}), dispersion modelling (Section~\ref{sec:lit-dispersion}), and exposure modelling (Section~\ref{sec:lit-exposure}). We then describe how the stages are coupled (Section~\ref{sec:lit-coupling}) and discuss what existing methods lack and how this work compares (Section~\ref{sec:lit-gap}).

\subsection{Traffic models}
\label{sec:lit-traffic}
The first stage models the movement of vehicles (location, speed, acceleration profiles). These traffic dynamics determine emissions; therefore, the choice of traffic models has significant effects throughout the pipeline. Traffic models are most commonly distinguished into three classes~\cite{madziel2023}. \textit{Macroscopic} models treat traffic as a fluid and work with aggregated speed and density on each link (e.g.\ VISUM~\citep{ptv_visum}), \textit{mesoscopic} models describe traffic per link through queue dynamics without resolving individual car-following, typically on one-second timesteps (e.g.\ MATSim~\citep{Horni2016}), and \textit{microscopic} models resolve full car-following and lane-changing and produce per-vehicle trajectories (SUMO~\citep{Lopez2018}, VISSIM~\citep{Fellendorf2010})~\cite{DIB2025104791, madziel2023}. Some studies also work with \textit{agent-based} models which add behavioural decisions such as mode choice and activity scheduling~\cite{HOFER20181,HULSMANN2014732, DIB2025104791}. For mesoscopic and microscopic work the most common tools are MATSim, VISSIM, and SUMO~\cite{Jang2025}. The studies compiled in the systematic review of \textcite{Jang2025} suggest a shift toward microscopic studies, and in particular towards SUMO, in recent years. These models are fed and calibrated from origin–destination matrices, traffic counts, surveys, camera data, and GPS traces~\cite{hirschmann2010,QUAASSDORFF2016}.

For an emission study, resolution is not a cosmetic choice. Microscopic models exhibit substantially more variability in emissions and even different total emissions than coarser models~\cite{Abou-Senna2013}. Congestion effects can be large: in a microsimulation of a Madrid roundabout, switching from free-flow to stop-and-go conditions on the same route increased emissions of nitrogen oxides (NO$_x$) and of particulate matter with an aerodynamic diameter of up to 10\,$\mu$m (PM$_{10}$) by 42.6\% and 27.9\% respectively, and across the whole modelled domain emissions roughly doubled (96.3\% for NO$_x$, 95.7\% for PM$_{10}$) between low- and high-congestion scenarios, coinciding spatially with stop-and-go zones that an average-speed description smooths away~\cite{QUAASSDORFF2016}. The price of this fidelity is computational cost: microscopic simulation becomes expensive at city or regional scale, which has motivated meso-to-micro upscaling~\cite{DIB2025104791,ZHOU2015} and the reuse of representative driving patterns computed once and applied broadly~\cite{LIU20111154}.

\subsection{Emission models}
\label{sec:lit-emission} 
The second stage derives pollutant mass emissions from vehicle dynamics. Traffic-related work concentrates on a recurring set of pollutants because of their particular health and regulatory relevance: nitrogen oxides (NO$_x$, in particular NO$_2$), \ac{PM}, in particular PM$_{10}$ and PM$_{2.5}$, carbon monoxide (CO), carbon dioxide (CO$_2$), and total hydrocarbons (THC)~\cite{madziel2023}. Of these, CO$_2$ is tracked mainly as a greenhouse gas rather than a local health concern~\cite{madziel2023}. Emissions divide further into primary pollutants, emitted directly, and secondary pollutants formed through atmospheric chemistry from primary precursors~\cite{samake2022}. Some studies broaden the inventory further, as \textcite{HUANG2009} do by comparing road-maintenance emissions with those of the disrupted traffic in a life-cycle assessment.

Emission models themselves come at two scales. \textit{Aggregate} or \textit{macroscopic} models such as COPERT~\cite{Ntziachristos2009}, HBEFA~\cite{hbefa_web}, and MOBILE~\cite{usepa2003mobile6} provide gram-per-kilometre factors as a function of average speed, vehicle class, and road type~\cite{madziel2023}. They are convenient for inventories but rest on dynamometer driving cycles that diverge from real driving~\cite{madziel2023}. \textit{Instantaneous} or \textit{microscopic} models instead evaluate each vehicle at roughly one-second resolution from its speed, acceleration, and road grade~\cite{Ma2012,madziel2023}; portable emission measurement systems (\acs{PEMS}) record such second-by-second emissions in real driving and are used to benchmark these models~\cite{Ma2012}. In their review of instantaneous models, \textcite{So21042018} split them into speed-based models such as VT-Micro~\citep{Rakha2004} and VERSIT+~\citep{Smit2007}, which fit statistical functions of speed and acceleration, and engine-based models such as CMEM~\citep{Barth2000} and \ac{PHEM}~\citep{Hausberger2009}, which compute engine power and engine speed explicitly and are therefore more reliable on roads with curvature and inclination. The two scales can be mixed: a microscopic traffic model can feed a macroscopic emission model~\cite{madziel2023}, but this must be done carefully, since \textcite{Abou-Senna2013} show that using average speeds rather than second-by-second operating-mode distributions overestimates total emissions and suppresses link-level spatial variation. Typical outputs are per-vehicle or per-link mass rates of NO$_x$, \ac{PM}, CO, CO$_2$, \ac{HC}, and fuel use, sometimes including cold-start and gear-shift effects~\cite{hirschmann2010,DIB2025104791}. A further complication is that brake, tire, and road-surface wear (so called non-exhaust pollutants) can also be substantial sources of emission~\cite{pant2013}.

This work uses the HBEFA4 emission model built into \ac{SUMO}, whose vehicle classes map directly onto the subsegments defined by HBEFA~\citep{sumo_hbefa4}. As with the other emission models in \ac{SUMO}, emission rates are evaluated from the vehicle's instantaneous speed and acceleration; for HBEFA4 these rates come from a polynomial curve fit to vehicle-category data~\citep{behrisch2024}. Resolution is in any case no guarantee of accuracy: across 50 validation studies \textcite{smit2010} report mean prediction errors within a factor of 2 for NO$_x$ and a factor of 3 for CO and \ac{PM}, and find no evidence that more complex models perform systematically better than simpler ones. For HBEFA in particular, \textcite{colberg2005} find that it overestimates NO$_x$ and CO relative to measurements in the Gubrist road tunnel in Switzerland, although this study predates HBEFA4 and tested an earlier version of the model; the authors note that differences in fleet composition could partly explain the deviation for light-duty vehicles. The emission models implemented in \ac{SUMO} output CO$_2$, CO, \ac{HC}, NO$_x$, PM$_x$, and fuel consumption~\citep{behrisch2024}.

\subsection{Dispersion models}
\label{sec:lit-dispersion}

The third stage calculates pollutant concentration fields (over space and time) from emissions. Two methodological families dominate. \textit{Statistical or data-driven} models, chiefly \ac{LUR}, estimate concentrations from spatial covariates~\cite{SONNENSCHEIN2025}. They are computationally cheap but correlation-based and struggle to isolate the traffic contribution from co-varying land use~\cite{SONNENSCHEIN2025}. Hybrid schemes that combine cellular automata with \ac{LUR} and calibrate the result by genetic algorithms can fold in meteorology and urban morphology at low computational cost~\cite{SONNENSCHEIN2025}. \textit{Deterministic, physics-based} models instead solve the advection--diffusion--reaction PDE for the concentration field, with terms like wind, eddy diffusivity, sources, chemistry, and deposition~\cite{samake2022}. These split into \textit{Eulerian} solvers that integrate the PDE on a fixed grid~\cite{Tirabassi2009} and \textit{Lagrangian} solvers that track virtual particles advected by the wind field~\cite{castelli}. The dispersion model used in this work belongs to the Eulerian family and is in spirit close to the PDE treatments of \textcite{samake2022} and \textcite{zhao2024}.

Concrete implementations span a wide fidelity range. Steady-state Gaussian models such as AERMOD~\cite{Cimorelli2005} and CAL3QHCR~\cite{Eckhoff1995} are often used in regulatory and academic contexts~\cite{Demiraslan2025}. Along a highway through complex terrain, \textcite{Demiraslan2025} found that AERMOD overpredicted near-road concentrations while CAL3QHCR agreed better with observed CO and NO$_x$, and that both performed poorly for PM$_{10}$. Street-canyon models such as \ac{OSPM}~\cite{Berkowicz2000} are fast but blind to complex geometries~\cite{HULSMANN2014732}, while regional chemical-transport models such as CAMx~\cite{ramboll2026camx} carry full atmospheric chemistry~\cite{PICCOLI2024}. High-resolution Lagrangian models (GRAL,~\cite{segura2025}) and CFD-grade solvers (VADIS,~\cite{SILVEIRA2023119481}; Fluidity,~\cite{WOODWARD2019}) offer the highest spatial resolution and physical detail at the highest computational cost. The source geometry also matters: roads are conventionally modelled as line sources, but \textcite{WOODWARD2019} show, for a low-wind case, that representing vehicles as moving bodies that displace the air changes near-road concentrations substantially compared with a line source. The general trade-off is that spatial resolution and physical fidelity scale with computational cost.

\subsection{Exposure models}
\label{sec:lit-exposure}

The final stage connects the concentration field with locations of people. Exposure is the product of concentration, contact time, and inhalation rate, not ambient concentration alone. \textcite{monn1998} measured Swiss home indoor NO$_2$ at roughly two-thirds of outdoor levels, while \textcite{zhao2024} simulate the inhaled dose for moving people in an urban district directly. Reported health endpoints include premature mortality~\cite{PICCOLI2024,SILVEIRA2023119481,wang2023} and years of life lost~\cite{PICCOLI2024,SILVEIRA2023119481}, short-term morbidity such as emergency doctor visits~\cite{ZHANG2013}, cancer risk from PM$_{2.5}$~\cite{zhao2024}, and monetary damages per day~\cite{SILVEIRA2023119481}. The \ac{WHO} Air Quality Guidelines serve as a reference for concentration–response functions in health impact assessment~\cite{PICCOLI2024,SILVEIRA2023119481}. A recurring simplification is that the exposed population is treated as static at its residential address~\cite{PICCOLI2024,ZHANG2013}, even though \textcite{denazelle2013} note that time spent travelling can contribute a disproportionate share of daily NO$_2$ intake. \textcite{HATZOPOULOU2010315} are a notable exception, tracking individuals through their daily activity schedule to derive person-level NO$_x$ exposure. 

Measurements provide a useful ground truth to validate exposure models, but fixed monitoring networks are sparse: \ac{NABEL} covers the entire country with only 16 stations~\citep{nabel2025}. \textcite{Demiraslan2025} report that independent field campaigns were not feasible at the scale of their study area.

\subsection{How the stages are coupled}
\label{sec:lit-coupling}

Depending on the stages included in an investigation, only certain questions can be answered. Table~\ref{tab:lit_overview} lists, for each reviewed work, which of the four stages it covers and whether it simulates traffic or relies on observed data. \textit{Decoupled} approaches use observed or aggregated traffic data combined with an emission model~\cite{PICCOLI2024,SILVEIRA2023119481,wang2023,DIAS2016253}. These are computationally cheaper and simpler to implement, but cannot represent counterfactual scenarios and inherit the spatiotemporal resolution of their input data. \textit{Coupled} approaches use vehicle trajectories or link conditions from traffic simulation to feed emission models. Various attempts with different granularity levels can be found in the existing literature: microscopic traffic with macroscopic emissions~\cite{HULSMANN2014732,HATZOPOULOU2010315}, microscopic traffic with microscopic emissions~\cite{QUAASSDORFF2016,Abou-Senna2013,hirschmann2010,So21042018,WOODWARD2019}, and mesoscopic traffic with microscopic post-processing~\cite{ZHOU2015,DIB2025104791}.

Real-world validation of these couplings remains limited. \textcite{hirschmann2010} carry out instrumented field measurements; \textcite{QUAASSDORFF2016} cross-validate a microscopic model against COPERT and find the largest divergence in high-congestion regimes; and \textcite{Ma2012} benchmark four microscopic models against data from \acp{PEMS}. Across this body of work, a few findings recur. Average-speed models diverge mostly from higher-resolution models under congestion, although not always in the same direction~\cite{Abou-Senna2013,QUAASSDORFF2016}, and congestion changes not only the amount but also the spatial pattern of emissions~\cite{QUAASSDORFF2016}. Cold-start and road-grade effects matter~\cite{hirschmann2010} and are often omitted, whether from macroscopic emission models~\cite{DIB2025104791} or from the microscopic traffic simulators that feed them~\cite{So21042018}.

\subsection{What is missing, and how this work compares}
\label{sec:lit-gap}

Reviewing the related literature, existing work distributes unevenly along the four stages. Many studies end at the emission level~\cite{Abou-Senna2013,hirschmann2010,HUANG2009,So21042018}; others add dispersion but feed it with coarse traffic input~\cite{PICCOLI2024,SILVEIRA2023119481,wang2023}; and only a few close the loop to exposure and health~\cite{PICCOLI2024,SILVEIRA2023119481,ZHANG2013,wang2023,zhao2024}, as summarised in Table~\ref{tab:lit_overview}.

Two gaps stand out. First, pipelines that couple traffic, emissions, deterministic dispersion, and exposure end-to-end are rare. \textcite{zhao2024} is the closest precedent and was used here as first evidence, but it addresses pedestrians in an urban district in Riverside, CA of about 1.8 by 0.9\,km (353 by 185 cells of 5\,m~\citep{zhao2024}), whereas the present work resolves resident exposure across six entire settlements. \textcite{HATZOPOULOU2010315} also span all four stages, but rely on the off-the-shelf CALPUFF~\citep{scire2000calpuff} Gaussian-puff model rather than developing their own dispersion stage. Second, the coupled studies that drive dispersion from observed or aggregated counts~\cite{PICCOLI2024,SILVEIRA2023119481,wang2023} cannot represent counterfactual traffic scenarios such as a partial ban or a rerouting share, because their traffic input is fixed.

These gaps motivate the method developed in Section~\ref{sec:methods} and its demonstration in the Gotthard case study in Section~\ref{sec:casestudy}: an end-to-end coupled pipeline that trades some of the dispersion fidelity of CFD-grade models for a computationally tractable workflow covering several settlements.

\begin{table}[H]
\centering
\caption{\textbf{Reviewed simulation pipelines by stage}. \checkmark{} indicates the work
performs the stage; $\sim$ a partial or simplified treatment. ``Sim.\ traffic''
marks works that simulate traffic rather than using observed/aggregated data.}
\label{tab:lit_overview}
\renewcommand{\arraystretch}{1.12}
\setlength{\tabcolsep}{6pt}
\small
\begin{tabular}{l c c c c}
\toprule
\textbf{Reference} & \textbf{Sim.\ traffic} & \textbf{Emissions} & \textbf{Dispersion} & \textbf{Exposure} \\
\midrule
\textcite{Abou-Senna2013}     & \checkmark & \checkmark & --         & --        \\
\textcite{castelli}           & --         & \checkmark & \checkmark & --        \\
\textcite{Demiraslan2025}     & --         & \checkmark & \checkmark & --        \\
\textcite{DIAS2016253}        & --         & \checkmark & \checkmark & --         \\
\textcite{DIB2025104791}      & \checkmark & \checkmark & --         & --        \\
\textcite{HATZOPOULOU2010315} & \checkmark & \checkmark & \checkmark & \checkmark \\
\textcite{hirschmann2010}     & \checkmark & \checkmark & --         & --        \\
\textcite{HOFER20181}         & \checkmark & \checkmark & --         & --        \\
\textcite{HUANG2009}          & \checkmark & \checkmark & --         & --        \\
\textcite{HULSMANN2014732}    & \checkmark & \checkmark & \checkmark & --        \\
\textcite{LIU20111154}        & --         & \checkmark & --         & --        \\
\textcite{Ma2012}             & --         & \checkmark & --         & --        \\
\textcite{PICCOLI2024}        & --         & \checkmark & \checkmark & \checkmark \\
\textcite{QUAASSDORFF2016}    & \checkmark & \checkmark & --         & --        \\
\textcite{samake2022}         & \checkmark & \checkmark & \checkmark & --        \\
\textcite{segura2025}         & \checkmark & \checkmark & \checkmark & --        \\
\textcite{SILVEIRA2023119481} & --         & \checkmark & \checkmark & \checkmark \\
\textcite{So21042018}         & \checkmark & \checkmark & --         & --        \\
\textcite{SONNENSCHEIN2025}   & --         & --         & \checkmark & --        \\
\textcite{wang2023}           & --         & \checkmark & \checkmark & \checkmark \\
\textcite{WOODWARD2019}       & \checkmark & \checkmark & \checkmark & $\sim$    \\
\textcite{ZHANG2013}          & --         & \checkmark & \checkmark & \checkmark \\
\textcite{zhao2024}           & \checkmark & \checkmark & \checkmark & \checkmark \\
\textcite{ZHOU2015}           & \checkmark & \checkmark & --         & --        \\
\midrule
\textbf{This work}            & \checkmark & \checkmark & \checkmark & \checkmark \\
\bottomrule
\end{tabular}
\end{table}

\newpage
\section{Methods}
\label{sec:methods}

This section describes the pipeline independently of any particular site: it defines what each stage does, not how the case-study data were sourced or how the solver is implemented, which are given in Appendix~\ref{app:technical}. Its application to the Gotthard corridor, including all site-specific data and scenario settings, follows in Section~\ref{sec:casestudy}.

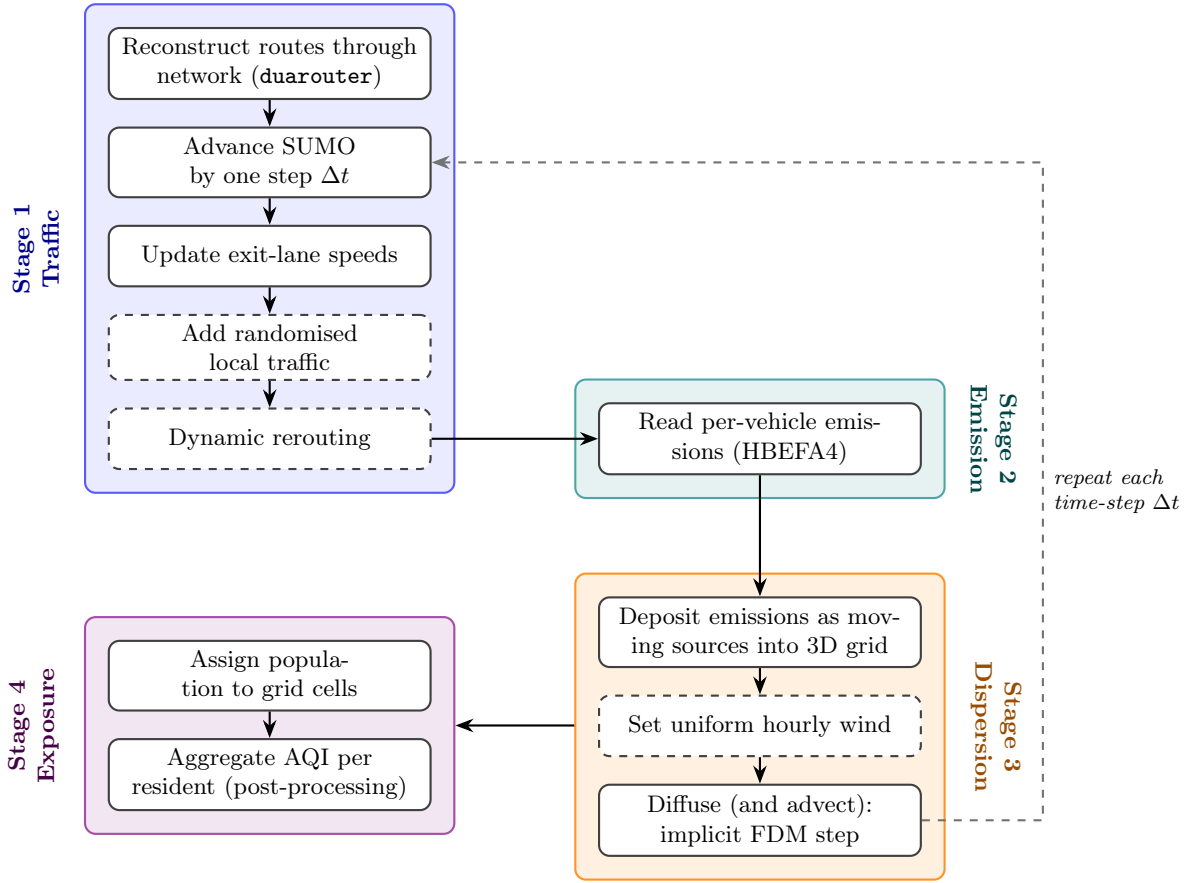
\begin{figure}[H]
\centering
\begin{tikzpicture}[
  node distance=3.5mm,
  pbox/.style={rectangle, rounded corners, draw=black!75, thick, fill=white,
               align=center, font=\footnotesize, text width=40mm, minimum height=8mm},
  obox/.style={pbox, dashed},
  arr/.style={-{Stealth[length=2.4mm]}, thick},
  looparr/.style={-{Stealth[length=2.4mm]}, thick, dashed, draw=black!55},
  stagelbl/.style={font=\bfseries\footnotesize, align=center}
]
% Stage 1 (top left)
\node[pbox] (route) {Reconstruct routes through network (\texttt{duarouter})};
\node[pbox, below=of route]   (advance) {Advance SUMO by one step $\Delta t$};
\node[pbox, below=of advance] (speed)   {Update exit-lane speeds};
\node[obox, below=of speed]   (rand)    {Add randomised local traffic};
\node[obox, below=of rand]    (reroute) {Dynamic rerouting};
% Stage 2 (top right)
\node[pbox, right=22mm of reroute] (emit) {Read per-vehicle emissions (HBEFA4)};
% Stage 3 (bottom right)
\node[pbox, below=16mm of emit] (deposit) {Deposit emissions as moving sources into 3D grid};
\node[obox, below=of deposit]   (wind)    {Set uniform hourly wind};
\node[pbox, below=of wind]      (diffuse) {Diffuse (and advect): implicit FDM step};
% Stage 4 (bottom left)
\node[pbox, left=22mm of deposit, yshift=-5.75mm] (assign) {Assign population to grid cells};
\node[pbox, below=of assign] (aqi) {Aggregate AQI per resident (post-processing)};

\draw[arr] (route)   -- (advance);
\draw[arr] (advance) -- (speed);
\draw[arr] (speed)   -- (rand);
\draw[arr] (rand)    -- (reroute);
\draw[arr] (deposit) -- (wind);
\draw[arr] (wind)    -- (diffuse);
\draw[arr] (assign)  -- (aqi);

\begin{scope}[on background layer]
  \node[draw=blue!65,   fill=blue!8,    rounded corners, thick, inner sep=3mm, fit=(route)(advance)(speed)(rand)(reroute)] (s1) {};
  \node[draw=teal!70,   fill=teal!10,   rounded corners, thick, inner sep=3mm, fit=(emit)] (s2) {};
  \node[draw=orange!85, fill=orange!12, rounded corners, thick, inner sep=3mm, fit=(deposit)(wind)(diffuse)] (s3) {};
  \node[draw=violet!70, fill=violet!10, rounded corners, thick, inner sep=3mm, fit=(assign)(aqi)] (s4) {};
\end{scope}

\draw[arr] (reroute.east) -- (emit.west);
\draw[arr] (emit.south) -- (deposit.north);
\draw[arr] (s3.west |- s4.east) -- (s4.east);
\draw[looparr] (diffuse.east) -- ++(16mm,0) |- (advance.east)
  node[pos=0.25, right, font=\scriptsize\itshape, align=left] {repeat each\\ time-step $\Delta t$};

\node[stagelbl, blue!55!black,   left=2mm of s1.west,  rotate=90,  anchor=south] {Stage 1\\Traffic};
\node[stagelbl, teal!55!black,   right=2mm of s2.east, rotate=-90, anchor=south] {Stage 2\\Emission};
\node[stagelbl, orange!60!black, right=2mm of s3.east, rotate=-90, anchor=south] {Stage 3\\Dispersion};
\node[stagelbl, violet!55!black, left=2mm of s4.west,  rotate=90,  anchor=south] {Stage 4\\Exposure};
\end{tikzpicture}
\caption{\textbf{Simulation pipeline (grouped by stage)}. The dashed arrow marks the per-time-step loop over traffic, emission, and dispersion. Dashed boxes mark optional modules: randomised local traffic, dynamic rerouting, and a uniform wind that adds advection to the diffusion step.}
\label{fig:pipeline}
\end{figure}

Following the four stages, the pipeline starts with traffic simulation that produces per-vehicle trajectories, continues with emission modelling that reads instantaneous rates from those trajectories, proceeds with a deterministic diffusion solver that transports the emitted mass through a three-dimensional concentration field, and concludes by aggregating the resulting concentrations into a health-relevant index for each resident. Figure~\ref{fig:pipeline} summarises the pipeline and maps its component steps to the four stages. The guiding design principle is a deliberate trade-off between physical fidelity and computational tractability. The pipeline resolves the full chain from moving vehicles to resident exposure for the residential area of a settlement, accepting simplifications in exchange for a method that runs end-to-end without specialised infrastructure such as large-scale compute clusters. The full pipeline is implemented in Python and is publicly available online: \url{https://github.com/cedriczeiter/zeiter_coupled_pipeline}.
 
\subsection{Traffic model}
\label{sec:meth-traffic}

The microscopic simulator \ac{SUMO} is used for traffic modelling. The road network of a study domain is constructed from \ac{OSM}~\citep{haklay2008osm} data and imported into SUMO's standard XML network format, preserving the spatial layout of streets, intersections, and speed limits from the actual infrastructure; topological errors in the \ac{OSM} data are corrected manually.

Traffic demand enters the pipeline as a set of vehicle routes through the domain, reconstructed from the available input data for the study at hand (Section~\ref{sec:casestudy-data}).

Within the simulation, vehicle positions and driving states are updated at every time-step and made accessible to external control via the \ac{TraCI} interface~\citep{wegener2008traci}, which lets a Python controller read vehicle states and inject or modify vehicles at runtime. Two optional modules extend the baseline demand through this interface: randomised local traffic, which adds trips made by residents within the domain that through-traffic records do not contain, and dynamic rerouting, which lets a share of vehicles recompute their route in response to simulated congestion. Both modules are described in Appendix~\ref{app:tech-modules}; their settings in the case study are given in Section~\ref{sec:casestudy-scenarios}.

\subsection{Emission model}
\label{sec:meth-emission}
For each simulation step, per-vehicle emission rates from \ac{SUMO}'s HBEFA4 emission model~\citep{sumo_hbefa4} are retrieved as follows: for every active vehicle, the current position and the CO, NO$_x$, and PM$_x$ emission rates (mass per time) are multiplied by the (time) step length to obtain the emitted masses. HBEFA4 polynomially models the emission rates as a function of the vehicle's emission and combustion class, and its instantaneous speed and acceleration~\citep{behrisch2024}. The emission class is a property of the \ac{SUMO} vehicle type; where a type carries no explicit HBEFA4 class, \ac{SUMO} applies its built-in default for the vehicle class. The fleet is therefore an input to the pipeline; the fleet used in the case study is reported in Appendix~\ref{app:tech-fleet}.

Which of the pollutants reported by HBEFA4 to carry forward is decided by the exposure metric rather than by the emission model. The \ac{AQI} defines concentration breakpoints for a fixed set of pollutants~\cite{usepa_aqi_2026}. Of these, CO, NO$_x$, and \ac{PM} are emitted directly by road traffic and are reported by HBEFA4, the latter two as total NO$_x$ and total PM$_x$, whereas CO$_2$ and \ac{HC} have no \ac{AQI} breakpoint and therefore cannot enter an \ac{AQI}-based exposure calculation at all.

The \ac{AQI} breakpoints are defined for NO$_2$ and for PM$_{2.5}$ and PM$_{10}$ separately. We identify the simulated NO$_x$ field with NO$_2$ and the simulated PM$_x$ field with PM$_{2.5}$, and report both under the latter names throughout the remainder of this paper. The two identifications rest on different footing. For \ac{PM} the mapping follows the established convention for road transport emission inventories: because the coarse fraction is negligible in vehicle exhaust, all exhaust \ac{PM} mass is taken to correspond to PM$_{2.5}$~\cite{emep2024}, which is why PM$_{2.5}$ rather than PM$_{10}$ is used. For nitrogen oxides the mapping is a simplification rather than a physical equivalence. Only a minority of tailpipe NO$_x$ leaves the exhaust as NO$_2$: monitoring across 43 sites in London gives a median primary NO$_2$ fraction of 10.6\,vol\,\%~\cite{carslaw2005}, while portable emission measurements on 39 Euro~6 diesel passenger cars give a study mean of 44\,\%~\cite{odriscoll2016}. The balance is \ac{NO}, and near-road measurements show that NO$_x$ comprises a higher fraction of \ac{NO} closer to the roadway than further away~\cite{Clements2009}. Regression of near-road monitoring data along an interstate highway gave on-road NO$_2$/NO$_x$ ratios of 0.25 to 0.35~\cite{RichmondBryant2017}, so at the road the simulated field may overstate NO$_2$ by a factor of about three to four.

\subsection{Dispersion model}
\label{sec:meth-dispersion}

\paragraph{Governing equation.} The concentration field $C(\mathbf{x},t)$ [mg\,m$^{-3}$] of each pollutant evolves according to a three-dimensional advection-diffusion-reaction \ac{PDE} with a moving vehicle source term~\cite{Tirabassi2009,samake2022}:
\begin{equation}
\frac{\partial C}{\partial t} + \nabla \cdot (\mathbf{v}\,C) = \frac{\partial}{\partial x}\Big(D_x\frac{\partial C}{\partial x}\Big) + \frac{\partial}{\partial y}\Big(D_y\frac{\partial C}{\partial y}\Big) + \frac{\partial}{\partial z}\Big(K_z(z)\frac{\partial C}{\partial z}\Big) - \lambda\,(C - C_{\mathrm{bg}}) + S(\mathbf{x},t),
\label{eq:pde}
\end{equation}
where $\mathbf{v}$ is the wind velocity field, $D_x, D_y$ [m$^2$\,s$^{-1}$] are constant horizontal turbulent diffusion coefficients, $K_z(z)$ [m$^2$\,s$^{-1}$] is a height-dependent vertical eddy diffusivity, $\lambda$ [s$^{-1}$] is a first-order loss rate acting on the excess over the ambient background concentration $C_{\mathrm{bg}}$, so that the background itself is not depleted, and $S$ is the vehicle emission source term [mg\,m$^{-3}$\,s$^{-1}$]. The loss term is applied only to PM$_{2.5}$, as a calibration surrogate for deposition and other removal processes the model does not represent explicitly (Section~\ref{sec:discussion}); $\lambda = 0$ for CO and NO$_2$. In the main model wind advection is omitted ($\mathbf{v} = \mathbf{0}$); Section~\ref{sec:windsens} activates $\mathbf{v}$, held spatially uniform over the domain and constant within each hour, to test the sensitivity of the results to this assumption.

In our study, the vertical diffusivity follows Monin--Obukhov similarity theory, in the same spirit as \textcite{zhao2024}:
\begin{equation}
K_z(z) = \frac{\kappa\, u_*\, z}{\phi_h(z/L)}, \qquad
\phi_h(\zeta) = \begin{cases}
0.74\,(1 - 9\zeta)^{-1/2}, & \zeta < 0 \text{ (unstable)} \\
0.74 + 4.7\,\zeta, & \zeta \ge 0 \text{ (stable)}
\end{cases}
\label{eq:kz}
\end{equation}
with von K\'arm\'an constant $\kappa = 0.35$~\citep{businger1971}, friction velocity $u_*$, and Obukhov length $L$. \textcite{zhao2024} derive these micrometeorological inputs with the AERMOD meteorological processor but report no formula for their own $K_z$, and their case study uses a fixed Obukhov length; we likewise use fixed scenario-level values of $u_*$ and $L$ rather than hourly meteorology, reported in Appendix~\ref{app:params}. No boundary-layer shape factor is applied, so Eq.~\eqref{eq:kz} is used up to the domain top even though Monin--Obukhov theory is strictly valid only within the surface layer; a shape factor such as $(1-z/z_i)^2$ would force $K_z=0$ exactly at the top Dirichlet face and was rejected for that reason (Appendix~\ref{app:tech-dispersion}).

\paragraph{Discretisation and boundary conditions.} Eq.~\eqref{eq:pde} is discretised with finite differences on a three-dimensional Cartesian grid that is uniform in the horizontal and geometrically stretched in the vertical, following \textcite{zhao2024}, so that resolution concentrates near the ground where vertical concentration gradients are steepest. At every time-step, the mass emitted by each vehicle is deposited as a moving source into the lowest-layer grid cell containing that vehicle. Time is advanced with an implicit backward-Euler scheme that is split into one-dimensional sweeps (Lie splitting), which is unconditionally stable, so the time-step is not restricted by the grid spacing and is instead set to match the one-second resolution of the traffic simulation. The lateral boundaries are held at the ambient background concentration by ghost cells one cell width beyond the domain, the top boundary is held at the background at the top face, and the ground is treated as impermeable, reflecting the absence of a surface sink. The discrete equations, the full boundary treatment, and implementation details of the case study are given in Appendix~\ref{app:tech-dispersion}.

\subsection{Exposure assessment}
\label{sec:meth-exposure}

\paragraph{Air quality metric.} Resident exposure is quantified using the \ac{AQI} of the \ac{EPA}~\citep{usepa_aqi_2026}, which maps a time-averaged concentration to a dimensionless 0--500 scale by piecewise-linear interpolation within predefined breakpoint bands:
\begin{equation}
I = I_{\mathrm{low}} + \frac{C - C_{\mathrm{low}}}{C_{\mathrm{high}} - C_{\mathrm{low}}} \cdot (I_{\mathrm{high}} - I_{\mathrm{low}}),
\label{eq:aqi}
\end{equation}
where $C$ is the time-averaged concentration and $C_{\mathrm{low}}, C_{\mathrm{high}}$ are the surrounding breakpoint concentrations with corresponding index values $I_{\mathrm{low}}, I_{\mathrm{high}}$. Following \ac{EPA} time-averaging conventions, CO is evaluated over an 8-hour window, NO$_2$ over a 1-hour window, and PM$_{2.5}$ over a 24-hour window. The evaluation height, the unit conversion, and the construction of the averaging windows from the simulation output are given in Appendix~\ref{app:tech-aqi}.

\paragraph{Combining pollutants.} The three sub-indices are summed without weighting to obtain a single value per resident. This departs from the \ac{EPA} convention, under which the \ac{AQI} is the maximum of its sub-indices, and is adopted because in low-traffic settings a single pollutant can dominate numerically through its background alone, as PM$_{2.5}$ does in the case study (Section~\ref{sec:res-villages}): taking the maximum would reduce the comparison to that pollutant, whereas summing preserves the contribution of all three. As a consequence, the summed value no longer carries the usual meaning of the \ac{AQI}: the \ac{EPA} scale is defined for one pollutant at a time, so the summed index serves only as a relative measure for comparing residents, settlements, and scenarios within this study.

\paragraph{Excess index.} Every sub-index is non-zero at the ambient background. The summed index therefore has a floor $I_{\mathrm{bg}}$, the index of the background concentrations, which is present in every cell and carries no information about traffic. We therefore report both the index $I$ itself, which is what the \ac{EPA} scale would assign a resident, and the excess index $\Delta I = I - I_{\mathrm{bg}}$, the part of a resident's score attributable to the simulated traffic. The \ac{AQI} maps concentration to index by linear interpolation within a series of fixed bands (Eq.~\ref{eq:aqi}), and each band has its own slope. When two residents' concentrations fall in the same band, an equal difference in $\Delta I$ means an equal difference in concentration. If a breakpoint separating two bands lies between them, this no longer holds: the same $\Delta I$ can correspond to different concentration differences on either side of that breakpoint. $\Delta I$ is therefore compared across settlements and dates as an index, on the common health-relevance scale the \ac{AQI} is designed to provide, rather than treated as a direct proxy for concentration.

\paragraph{Aggregation.} Individual \ac{AQI} scores are aggregated per settlement in three complementary ways, computed over every resident of the population falling within the domain; residents who share a grid cell each contribute that cell's \ac{AQI} separately, so a household of several people is counted once per person. The raw sum over all residents captures the total burden and scales with settlement size. The mean raw index per resident normalises by population and enables comparisons between settlements of different sizes. The mean excess index $\Delta I$ per resident removes the ambient floor and isolates the traffic-attributable part of that burden. The same three aggregations can also be computed for each pollutant individually, each against its own background floor.

\section{Case Study: Gotthard Corridor, Canton of Uri}
\label{sec:casestudy}

The value of this pipeline is showcased at the Gotthard transit corridor in the Swiss canton of Uri, which is especially interesting to study for two reasons: First, the corridor is one of the major traffic bottlenecks in Switzerland, and subject to political dispute about congestion-driven detour traffic~\citep{Campax2022, UrnerZeitung2023}, giving the assessment direct policy relevance. Second, the combination of a rural resident population with heavy seasonal transit congestion places the pipeline in a setting distinct from the urban case studies that dominate the coupled-pipeline literature. To the best of our knowledge, the corridor has not been modelled with a coupled traffic-to-exposure simulation.  This section collects everything that is specific to the site: the study area, the input data, the parameter values, and the simulated scenarios.

\subsection{Study area}
\label{sec:casestudy-area}

The study area is the main valley of Uri in central Switzerland, which runs from north to south between the southern arm of Lake Lucerne and the northern portal of the Gotthard road tunnel. The valley floor is narrow and its flanks are steep, so that settlement and infrastructure share the same strip of flat ground. Over long stretches the valley bottom is fully occupied by the A2 motorway, the cantonal road, and the railway line.

The Gotthard road tunnel, whose northern portal lies at Göschenen at the southern end of the study area, consists of a single bore. For safety reasons, inflow into the tunnel is metered to a maximum of 1000 vehicle units per hour and direction, with a heavy goods vehicle counting as three units~\citep{ASTRA202120Jahre}. This metering caps A2 throughput at the portal independently of demand: demand above the cap queues on the A2, and part of it diverts onto the parallel cantonal road.

Between Silenen and Göschenen only two roads follow the valley: the A2 and the cantonal road. The cantonal road is therefore the only route within the valley available for vehicles that leave the queuing A2, and it is a normal, undivided road built for local rather than transit volumes. It is well developed as far as Amsteg, but narrows markedly further south; in the stretch between Silenen and Wassen, where the valley is narrowest, any deviation from the motorway leads directly through village centres, and tight curves make passing oncoming vehicles difficult.
 
A categorically different alternative to the tunnel is the Gotthard Pass road, which crosses the Alps at a separate point and is closed to all traffic between Hospental and Airolo for about half of the year; in the most recently documented winter it was closed from 7 November 2025 to 8 May 2026~\citep{astra_2025_gotthard_close, astra_2026_gotthard_open}. Because vehicles bound for the pass also reach it via the cantonal road through the valley, an open pass adds pass-bound traffic to the cantonal road on top of any tunnel-avoiding traffic, rather than diverting demand away from the villages.
 
\begin{figure}[h]
\centering
\includegraphics[width=\linewidth]{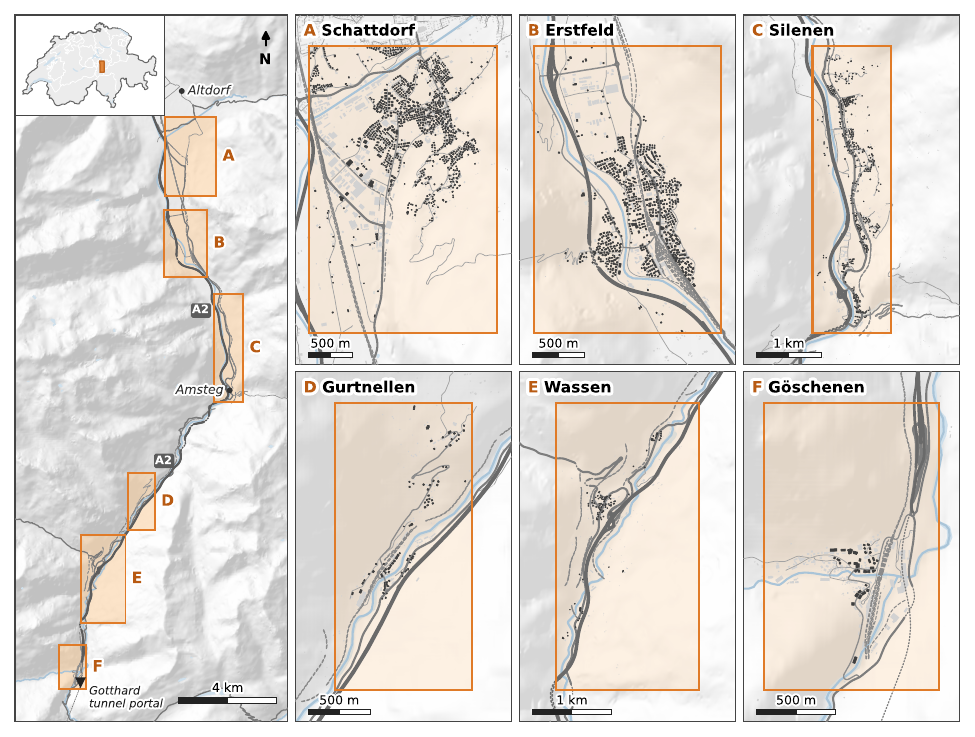}
\caption[Study area and simulation domains]{\textbf{Study area and simulation domains.} \textit{Left}: the main valley of Uri from Altdorf to the northern portal of the Gotthard road tunnel, with the six simulation domains (orange, A--F) along the A2 and the cantonal road; the inset locates the map within Switzerland. \textit{Right}: one panel per domain, ordered from north to south. Dark grey polygons are the residential buildings matched to SynPop residents, light grey polygons other buildings, grey lines roads, dashed lines the railway and blue lines rivers.}
\label{fig:all_villages_squares}
\end{figure}

Six villages along this corridor are investigated: Schattdorf, Erstfeld, Silenen, Gurtnellen, Wassen, and Göschenen. Only the residential areas of each village are included in the simulation domain, rather than the full administrative extent, both to focus the assessment on the inhabited built environment and to keep computational domains tractable (Figure~\ref{fig:all_villages_squares}).
 
\subsection{Traffic, fleet, and population data}
\label{sec:casestudy-data}

\paragraph{Traffic data.} The through-traffic demand is taken as given from a \ac{SUMO} simulation of the corridor produced for an unpublished master's thesis (see Acknowledgements); its vehicle trajectories through each village serve as input, and the demand reconstruction itself is not described here. Two dates were generated on request for this study: 20 July 2024, which recorded one of the highest southbound volumes towards the tunnel and pass among the days examined, and 20 November 2024, which recorded one of the lowest measured volumes in 2024 and serves as the low-traffic reference. Vehicles without a matched entry-exit pair are discarded, since they most likely represent residents rather than through-traffic. Because the underlying demand model explains detector counts with as little non-transit traffic as possible, local traffic is modelled separately (Section~\ref{sec:casestudy-scenarios}).

\paragraph{Fleet.} The vehicle fleet is inherited from the same demand model and comprises eight through-traffic vehicle types, plus one type for randomised local traffic. None of them carries an explicit HBEFA4 emission class, so each emits according to the \ac{SUMO} default class for its vehicle class (Section~\ref{sec:meth-emission}). The fleet is detailed in Appendix~\ref{app:tech-fleet}, and Section~\ref{sec:discussion} discusses the consequences of this simplification.

\paragraph{Population data.} Resident locations are taken from the Swiss Synthetic Population (SynPop 2022), a dataset of 8\,950\,435 individuals~\cite{ARE2024SynPop}. Home coordinates are anonymised by random shuffling within clusters of at least 1500 persons~\cite{ARE2024SynPop}. Records are filtered to the Uri canton bounding box, converted from the Swiss LV95 coordinate system to WGS84 and then to \ac{SUMO} network coordinates, and matched to the nearest grid cell of the concentration domain. Building footprints are queried from \ac{OSM} via the Overpass API and cached per village. They are used for visualisation, and the house-level statistics of Section~\ref{sec:results} are computed over the buildings returned by reverse geocoding of the SynPop coordinates, filtered by \ac{OSM} building tag. The driving-licence, car-ownership, and employment attributes of the SynPop records additionally parametrise the randomised local-traffic module, as described in Appendix~\ref{app:tech-modules}.

\subsection{Physical parameters and background}
\label{sec:casestudy-params}

Turbulent diffusion coefficients and ambient background concentrations for each pollutant were chosen heuristically, since no measurements of traffic plumes were available for the corridor. The selection procedure and the values adopted are reported in Appendix~\ref{app:params}, together with the Monin--Obukhov inputs of Eq.~\eqref{eq:kz} and the PM$_{2.5}$ loss rate. The background concentration of each pollutant maps, through the \ac{EPA} breakpoints of Eq.~\eqref{eq:aqi}, to a fixed sub-index that is present in every cell of the six domains and carries no information about traffic; summed over the three pollutants, this floor is the quantity $I_{\mathrm{bg}}$ of Section~\ref{sec:meth-exposure} and is subtracted to obtain the excess index $\Delta I$ reported in Section~\ref{sec:res-villages}.

\subsection{Simulation scenarios}
\label{sec:casestudy-scenarios}

Four scenarios are simulated on 20 July 2024, and Scenario~C is additionally simulated on 20 November 2024. Scenario~A simulates detour traffic only, with no local traffic and no rerouting; it appears as the left-hand panel in Figure~\ref{fig:random}. Scenario~B1 adds randomised local traffic with the weekend spawn profile, and Scenario~B2 adds randomised local traffic with the weekday spawn profile (Appendix~\ref{app:tech-modules}); neither enables rerouting. Both are shown here for 20 July 2024 in Figure~\ref{fig:random} together with Scenario~A; Scenario~B2 is also used alone in Figure~\ref{fig:timeofday}. Scenario~C adds randomised local traffic with the spawn profile appropriate to the calendar date, that is, the weekend profile on 20 July 2024 (a Saturday) and the weekday profile on 20 November 2024 (a Wednesday), and additionally enables dynamic rerouting at 5\,\% of vehicles per six-minute interval. Scenario~C applies this share to all six villages for consistency, although in villages dominated by a single road rerouted vehicles have few alternative paths. Scenario~C is used in Figure~\ref{fig:erstfeld_reduced} and for all cross-village \ac{AQI} comparisons in Section~\ref{sec:res-villages}.

A further sweep of the rerouting share, 0, 5, 10, and 15\,\% while keeping the same local-traffic profile as Scenario~C, is used in Figure~\ref{fig:reroute} to examine the rerouting effect on its own; the 0 and 5\,\% points coincide with Scenarios~B1 and~C in Schattdorf, and only the 5\,\% value is carried forward as Scenario~C for the remaining comparisons.

\section{Results}
\label{sec:results}

The results below are drawn from the Gotthard case study and serve to illustrate what the pipeline can resolve, rather than standing as a general claim about traffic-related exposure beyond this corridor. Section~\ref{sec:discussion} returns to the boundary between the two. 

\begin{figure}[H]
    \centering
    \includegraphics[width=0.9\linewidth]{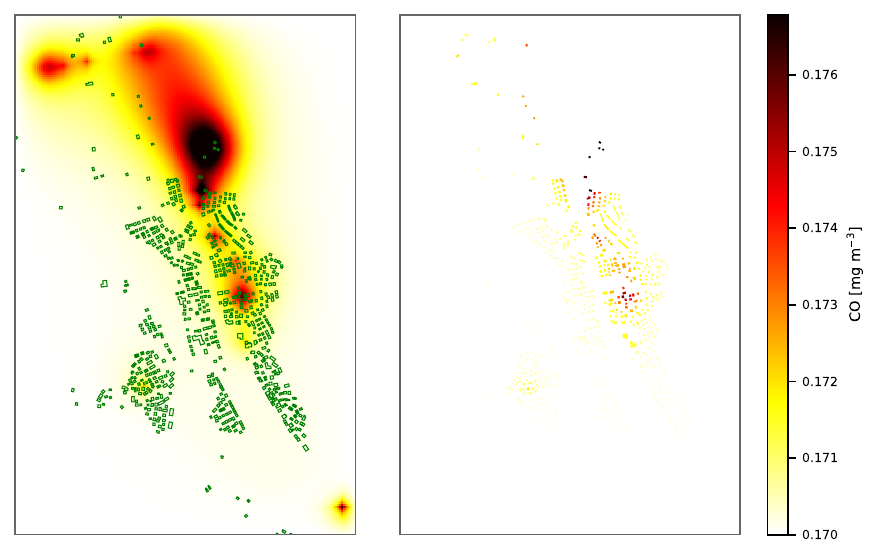}
    \caption{\textbf{Reduction of the CO concentration field to residential buildings} (Erstfeld, Scenario~C, 20~July, 6\,PM). \textit{Left}: full field, with house outlines in green; the road corridor saturates the colour scale. \textit{Right}: the same data with each building polygon coloured by the concentration at its grid cell. Both panels share one colour scale.}
    \label{fig:erstfeld_reduced}
\end{figure}

Both dates were simulated over full days: the high-traffic day of 20 July 2024 and the low-traffic reference day of 20 November 2024. We first isolate the individual model effects on the high-traffic day, namely time of day (Section~\ref{sec:res-time}), randomised local traffic (Section~\ref{sec:res-random}) and dynamic rerouting (Section~\ref{sec:res-reroute}), then test the no-wind assumption (Section~\ref{sec:windsens}), and finally compare all six villages on both dates under Scenario~C (Section~\ref{sec:res-villages}). The simulation saved hourly grid snapshots, which were rendered as heat maps; a representative set covering all villages is provided in Appendix~\ref{app:heatmaps}. Because most of the emitted mass never reaches a dwelling, concentrations are reported at the grid cells occupied by buildings matched to SynPop residents (Section~\ref{sec:casestudy-data}) rather than over the full field. Figure~\ref{fig:erstfeld_reduced} contrasts the two views for Erstfeld: the full field is dominated by the corridor north of the residential area, which saturates the colour scale, whereas the houses sit only barely above background. Figures~\ref{fig:random} and~\ref{fig:reroute} use the house-only view throughout to keep the comparison focused on residents rather than on the road itself. Sections~\ref{sec:res-time} to~\ref{sec:res-reroute} use single-hour snapshots of house-level mean concentrations, with the hour chosen because it shows the effect most clearly, so the reported patterns are illustrative and not typical of every hour.

\subsection{Temporal variation}
\label{sec:res-time}

\begin{figure}[H]
\centering
\includegraphics[width=\linewidth]{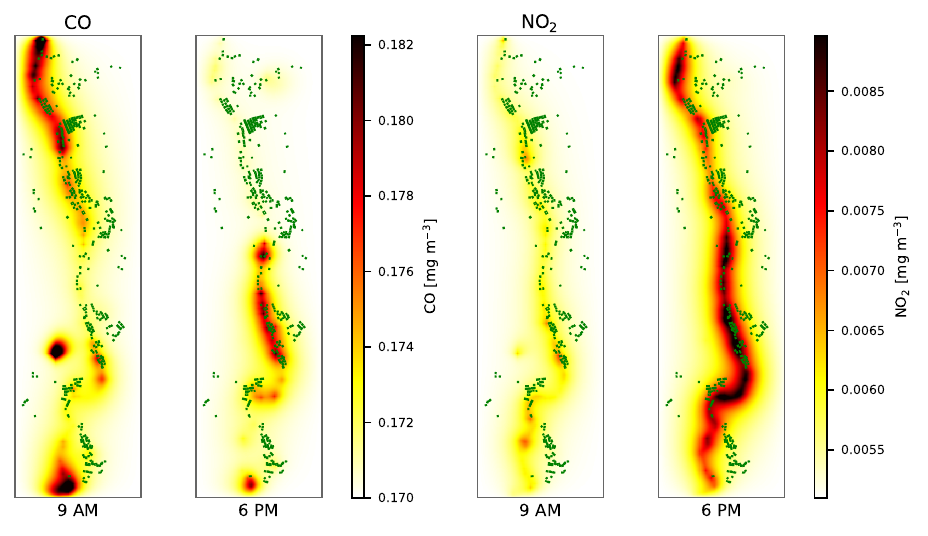}
\caption{\textbf{Temporal variation of CO (left pair) and NO$_2$ (right pair)} (Silenen, at 9\,AM and 6\,PM, Scenario~B2, 20~July). Colour scales are shared within each pair but differ between pollutants, so the two pairs are not directly comparable to one another. Each scale is capped so that the lower concentrations reaching residential buildings, the values of interest here, remain visible; the saturated colour along the road is therefore illustrative of relative change rather than a readout of the true peak. Green outlines mark residential buildings. CO peaks in the morning, NO$_2$ in the evening.}
\label{fig:timeofday}
\end{figure}

Because the simulations are driven by historical traffic that varies through the day, concentrations show pronounced temporal structure. Figure~\ref{fig:timeofday} shows Silenen in the morning and the evening on the high-traffic day, and the two pollutants peak at different times. CO is highest in the morning, with hotspots at the northern and southern ends of the road, and has fallen back towards background along most of the village by evening, although the house-level mean itself barely moves. NO$_2$ behaves in the opposite way: by evening an elevated band follows the entire road through the village, directly beside the houses, and the house-level mean rises noticeably. A change in vehicle numbers alone would move both pollutants in the same direction, so the divergence points to a change in the composition or driving state of the traffic, for example the share of light delivery vehicles and trucks, rather than to vehicle numbers alone. Because the health relevance of a pollutant depends on the duration of exposure as well as its magnitude, the AQI breakpoints used here are defined on time-averaged rather than instantaneous concentrations (Section~\ref{sec:meth-exposure}). Snapshots such as Figure~\ref{fig:timeofday} therefore serve only to show the temporal structure, and the cross-village exposure comparison below is reported through the time-averaged \ac{AQI}.

\subsection{Effect of randomised local traffic}
\label{sec:res-random}
 
\begin{figure}[H]
\centering
\includegraphics[width=\linewidth]{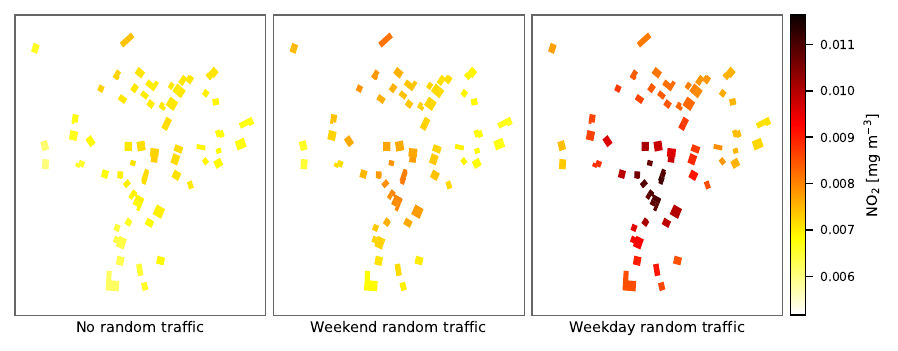}
\caption{\textbf{Effect of randomised local traffic} on residential NO$_2$ exposure (Wassen, 20~July, 6\,PM), house-only view on a shared colour scale. \textit{Left}: Scenario~A, detour traffic only. \textit{Centre}: Scenario~B1, weekend spawn profile. \textit{Right}: Scenario~B2, weekday spawn profile.}
\label{fig:random}
\end{figure}

The effect of randomised local traffic on residential NO$_2$ depends strongly on the village and on the spawn profile. Figure~\ref{fig:random} shows the clearest case, Wassen: relative to detour traffic alone, house-level NO$_2$ increases slightly under the weekend profile and markedly under the weekday profile, most strongly in the centre of the village. Across the six villages, the weekday profile, with its evening commuter peak, adds at least as much NO$_2$ as the weekend profile everywhere, and the effect is largest in Wassen rather than in the most populous village, Schattdorf, so the effect does not simply follow population. In a couple of villages, local traffic slightly lowers residential NO$_2$ at this hour instead; a plausible reason is that the added vehicles change when and where through traffic queues, but this was not investigated. Whether these effects generalise beyond the six villages studied is unclear, because the spawn parameters were set heuristically rather than from historical local-traffic data (Appendix~\ref{app:tech-modules}). Since randomised traffic represents the regular movement of residents, these vehicles are not themselves part of the detour traffic under study; their contribution to congestion nonetheless affects the detoured vehicles and is therefore retained in Scenario~C.

\subsection{Effect of dynamic rerouting}
\label{sec:res-reroute}
 
\begin{figure}[H]
\centering
\includegraphics[width=0.88\linewidth]{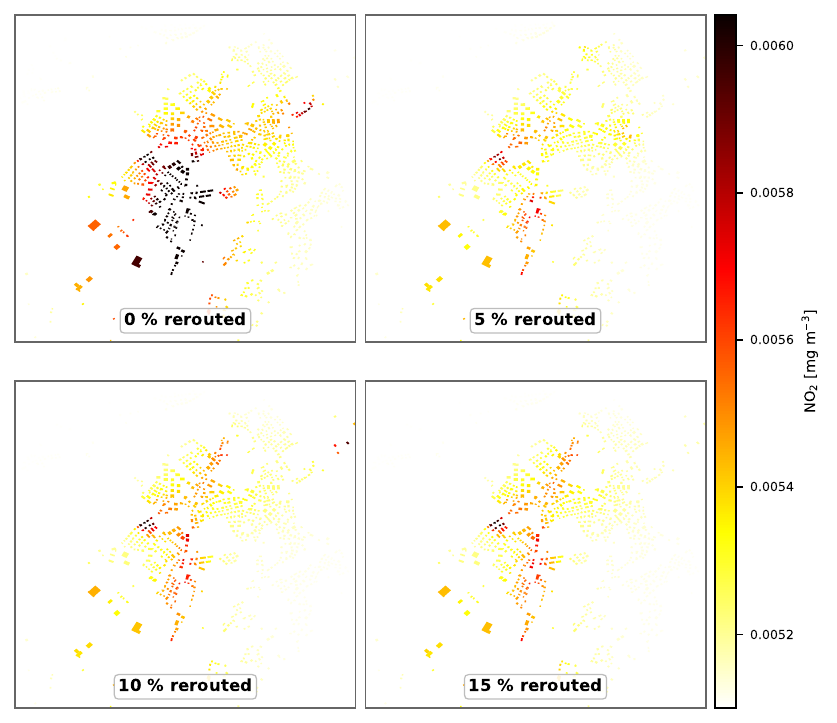}
\caption{\textbf{Effect of dynamic rerouting} on residential NO$_2$ exposure (Schattdorf, 20~July, 8\,PM, weekend spawn profile), house-only view on a shared colour scale. Panels show 0\,\%, 5\,\%, 10\,\% and 15\,\% of vehicles rerouted per 6-minute interval; the 0\,\% panel corresponds to Scenario~B1 and the 5\,\% panel to Scenario~C. NO$_2$ falls from 0 to 5\,\% and does not change further at higher shares.}
\label{fig:reroute}
\end{figure}

In Schattdorf, residential NO$_2$ is highest without rerouting and lower with it (Figure~\ref{fig:reroute}). Without rerouting, the highest values occur in a cluster of streets in the south-western part of the village; with rerouting, this cluster largely disappears and the remaining elevated values follow only a few streets. CO barely responds. Most of the reduction is reached already at the lowest share tested, and increasing the share further changes little: once the vehicles with a faster alternative inside the domain have switched, selecting more vehicles for rerouting finds few further route changes to make. The result comes from a single run in a single village. This effect was investigated for Schattdorf specifically because its network offers residents multiple parallel routes; villages dominated by a single main road have little for rerouting to act on.
 
\subsection{Wind sensitivity}
\label{sec:windsens}

\begin{figure}[H]
\centering
\includegraphics[width=0.85\linewidth]{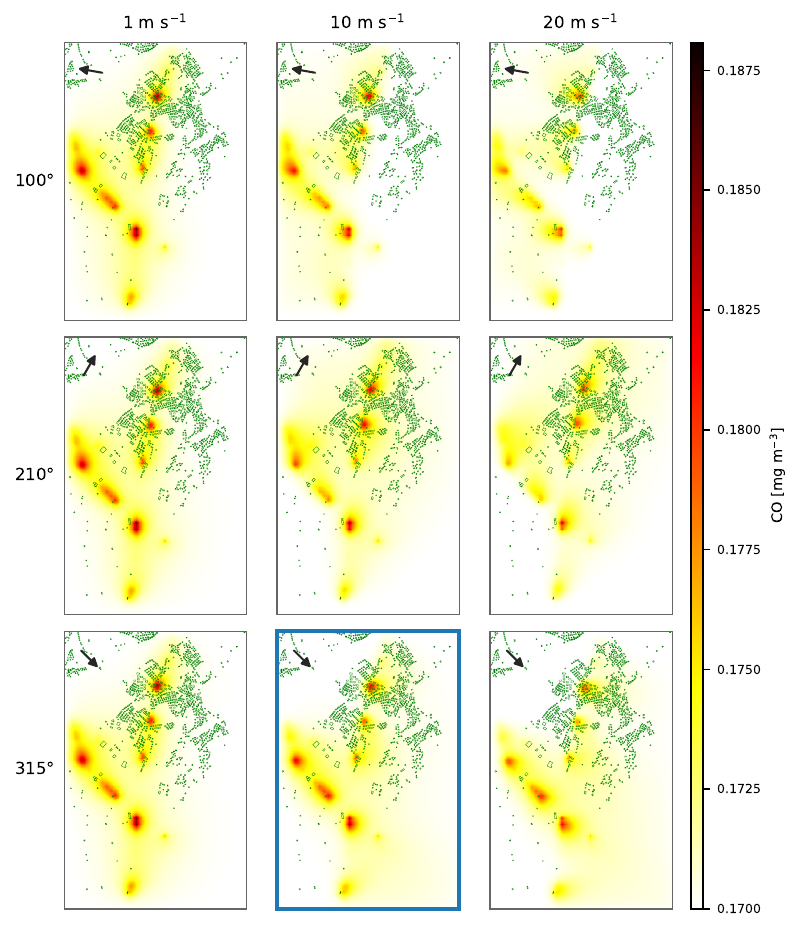}
\caption{\textbf{CO concentration in Schattdorf} at 9\,AM on 20~July (Scenario~C) under a spatially uniform wind, swept over three directions (rows, compass bearing the wind blows from: 100\textdegree, 210\textdegree, 315\textdegree) and three speeds (columns: 1, 10, 20\,m\,s$^{-1}$). Arrows show the direction in which the wind blows. All panels share one colour scale. Green outlines mark residential buildings. The framed panel at 315\textdegree\ and 10\,m\,s$^{-1}$ matches the wind used by \textcite{zhao2024}.}

\label{fig:windsens}
\end{figure}

The main pipeline omits wind (Section~\ref{sec:meth-dispersion}). The solver optionally accepts a spatially uniform wind, held constant within each hour (Appendix~\ref{app:tech-dispersion}), and Figure~\ref{fig:windsens} illustrates this capability for Schattdorf under Scenario~C, sweeping wind direction and speed for CO. Across the full sweep, the overall pattern of high and low concentration is set by the road and the built-up area rather than by the wind: the same hotspots appear in every panel, at every speed and every direction tested. What the wind changes is more subtle: increasing the speed smears these hotspots out somewhat and softens their peak, while rotating the direction mainly stretches the plume slightly further to one side of the road rather than shifting it to a different part of the village. This sweep is meant to demonstrate that the solver can incorporate wind, not to represent the actual valley wind field, which is shaped by the orography (Section~\ref{sec:discussion}).

\subsection{Cross-village exposure}
\label{sec:res-villages}
 
\begin{figure}[H]
\centering
\includegraphics[width=\linewidth]{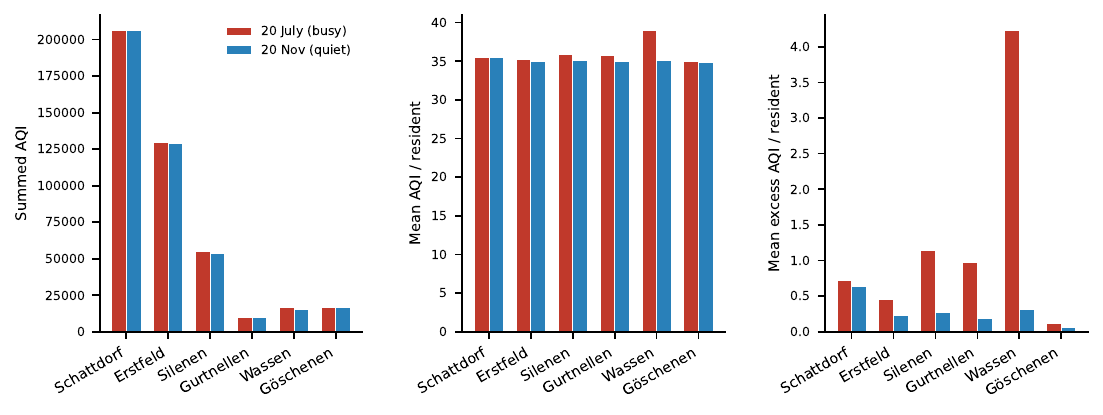}
\caption{\textbf{Cross-village comparison} of resident AQI exposure under Scenario~C for the high-traffic day (20~July, red) and the low-traffic reference day (20~November, blue). \textit{Left}: unweighted sum of the raw index over all residents, which scales with village population. \textit{Centre}: mean raw index per resident. \textit{Right}: mean excess index per resident, $\Delta I$, with the fixed ambient floor of 34.8 points removed; this is the quantity that isolates the traffic-attributable burden. Values are summed CO, NO$_2$ and PM$_{2.5}$ sub-indices.}
\label{fig:aqi_summary}
\end{figure}

All six villages were evaluated on both dates under Scenario~C, the best-estimate configuration with 5\,\% rerouting and calendar-appropriate randomised traffic. Figure~\ref{fig:aqi_summary} reports the raw index; the comparisons below are stated in terms of the excess index $\Delta I$ of Section~\ref{sec:meth-exposure}, because the ambient floor (Appendix~\ref{app:params}) dominates the raw value in every village and date and would otherwise swamp every comparison. Raw values are given alongside so that the two can be related.

\paragraph{Total and mean exposure.} Summed over all residents, the raw index simply tracks population size, as it must: the largest village carries by far the largest total burden, and the smallest villages the least, on both the busy and the quiet day. Per resident, the traffic-attributable part tells a different story. On the high-traffic day, Wassen's mean excess is markedly higher than that of any other village, followed by Silenen and Gurtnellen, ahead of Schattdorf, Erstfeld and Göschenen. On the quiet day the order changes: Schattdorf now has the highest mean excess, roughly double Wassen's, with the other villages trailing well behind. The ratio between the two days separates the villages into groups. Where diverted traffic passes close to dwellings, the busy day raises the traffic-attributable burden substantially in Wassen, Gurtnellen and Silenen. In Schattdorf the burden is almost the same on both days: the quiet day is a Wednesday with the weekday local-traffic profile, and residents' own trips take the place of the detour traffic of the busy Saturday. Erstfeld and Göschenen remain low on both days because most of their dwellings lie away from the corridor (Appendix~\ref{app:heatmaps}).

\paragraph{Which pollutant drives exposure.} PM$_{2.5}$ has the largest raw sub-index in every village-date combination, which is why the sub-indices are summed rather than reduced to their regulatory maximum: the maximum would report little more than the PM$_{2.5}$ background. That dominance is almost entirely background. Of the excess, NO$_2$ is the largest contributor in most villages, with PM$_{2.5}$ and CO playing a visibly smaller role. PM$_{2.5}$ reaches its largest share in Wassen, Erstfeld and Gurtnellen and stays negligible in Göschenen on the busy day, where its narrow plume stays on the road east of the dwellings (Appendix~\ref{app:heatmaps}). This attribution does not survive a more realistic treatment of NO$_2$: if only a modest fraction of the traffic-attributable NO$_x$ is counted as NO$_2$, within the measured range of on-road NO$_2$/NO$_x$ ratios~\citep{RichmondBryant2017}, PM$_{2.5}$ becomes the largest contributor in most villages instead. The ranking of villages by mean excess, in contrast, is unchanged on both dates. The pollutant attribution also depends on the fleet, on the pollutant-specific calibration targets (Appendix~\ref{app:params}), and on the AQI breakpoints, so it illustrates what the pipeline can resolve and is not a quantitative finding about the corridor (Section~\ref{sec:disc-implications}).

\paragraph{Summary.} The traffic-attributable burden per resident is highest where diverted traffic passes close to dwellings: in Wassen, where the motorway exit discharges into the village, and in Silenen and Gurtnellen, where the cantonal road runs through the built-up area. In these villages the busy day raises the burden substantially. Schattdorf, the largest of the six villages, carries by far the largest total burden, but its per-resident burden is set mainly by residents' own trips and hardly differs between the two days. This ordering of villages holds whether the full NO$_x$ field or only a smaller fraction of it is counted as NO$_2$; which pollutant dominates does not. Reporting the raw index conceals essentially all of this: on the quiet day the six raw means are nearly indistinguishable from each other, even though the traffic-attributable burden on that same day differs strongly across the same six villages.
 
\section{Discussion}
\label{sec:discussion}
 
\subsection{What the coupled pipeline enables}
\label{sec:disc-position}
 
The pipeline was built to close the gaps identified in Section~\ref{sec:lit-gap}: it spans all four stages, retains vehicle-level traffic dynamics, covers full days, and runs for the residential areas of entire villages. The case study shows that this coupling yields information that no single stage provides on its own. First, because the raw \ac{AQI} in a low-background setting is dominated by an ambient floor, differences between villages and days only become visible once the index is split into background and traffic-attributable excess. This decomposition does not depend on the corridor and can be applied to other index-based exposure assessments in clean-background regions. Second, evaluating the concentration field at dwellings rather than along roads orders settlements by where traffic passes relative to where people live, rather than by population, and separates villages whose burden follows the detour traffic from a village whose burden follows its residents' own trips. Third, because traffic is simulated rather than prescribed, traffic scenarios can in principle be represented before any intervention is implemented, although no real intervention was simulated here.
 
Relative to the closest prior work, the contribution lies in coverage and reproducibility rather than in physical fidelity. \textcite{zhao2024} couple \ac{SUMO} to a grid-based Eulerian dispersion model for pedestrians in an urban district, and \textcite{HATZOPOULOU2010315} link an activity-based demand model to the CALPUFF Gaussian puff model at city scale. The present pipeline extends the first to resident exposure across six settlements over full days and, unlike the second, keeps every stage within a single open implementation. To our knowledge, none of these works releases its full integrated pipeline.

\subsection{Limitations by stage}
\label{sec:disc-limitations}

The limitations are organised along the four stages of Figure~\ref{fig:fourstage} and described here as properties of the method. Their consequences for the case-study results are collected once, in Section~\ref{sec:disc-implications}.

\paragraph{Traffic.} The through-traffic demand and its vehicle types, including their car-following parameters, are inherited from the corridor simulation described above; as with any input taken from data we were given, its quality is not something we control. Because the HBEFA4 model responds to the instantaneous vehicle state, smoothed driving tends to suppress the very emission peaks that motivate a microscopic coupling. Treating each village as an isolated domain adds two constraints. Congestion at the domain exit is imposed from the corridor simulation rather than emerging, so detour chains across several villages are not represented, and rerouted vehicles can only choose among paths inside a single village. Vehicles without a matched entry and exit are discarded, and inflows from minor roads were not available. The two optional mechanisms rest on assumptions rather than observations. Local-traffic spawn rates were derived heuristically from SynPop attributes that are themselves imputed by statistical behavioural models~\citep{ARE2024SynPop}, and the rerouting share was chosen for the visual plausibility of the resulting heat maps rather than calibrated against observed driver behaviour (Appendix~\ref{app:tech-modules}). Finally, the network imported from \ac{OSM} does not encode narrow passages where oncoming vehicles cannot pass, and simulated drivers obey every restriction encoded in the network.

\paragraph{Emissions.} Road slope is not represented because the network imported from \ac{OSM} carries no elevation, and cold-start effects are not incorporated. As detailed in Appendix~\ref{app:tech-fleet}, no emission class is assigned explicitly, and the fleet carries no representation of the age, fuel, or Euro-norm distribution of the vehicles in the corridor. The sign of the resulting error is unknown for this corridor: real-world NO$_x$ emissions of Euro~6 diesel cars alone ranged from 1 to 22 times the type-approval limit~\citep{odriscoll2016}, and emission factors derived in a Swiss road tunnel differed between weekdays and weekends, probably because of fleet composition~\citep{colberg2005}. Across validation studies in general, emission models have tended to overestimate NO$_x$~\citep{smit2010}. The identification of the NO$_x$ field with NO$_2$ introduces a second, deliberate bias. Only part of tailpipe NO$_x$ is emitted directly as NO$_2$: the median primary fraction across 43 London roadside sites was 10.6\,vol\,\%~\citep{carslaw2005} and the mean for 39 Euro~6 diesel cars was 44\,\%~\citep{odriscoll2016}, and the fraction varies with fuel type, vehicle technology, and operating conditions~\citep{carslaw2005}. These diesel figures are not representative of the petrol passenger cars that make up most of the simulated fleet. The remainder is \ac{NO}, which forms NO$_2$ by reaction with ozone. That reaction typically takes tens of seconds, but how much \ac{NO} is converted close to the road is limited by how quickly the plume mixes with ambient air~\citep{carslaw2005}. Finally, the model contains no chemistry and cannot track this conversion. Because NO$_2$ cannot exceed total NO$_x$, the simulated NO$_2$ field is an upper bound only relative to the simulated NO$_x$, whose own error has unknown sign. Similar caveats apply to \ac{PM} and CO, whose emission factors carry the same fleet uncertainty.

\paragraph{Dispersion.} The dispersion stage carries the strongest physical simplifications: wind is omitted and the ground is flat. In mountain valleys, mean flow and turbulence are heavily modified by the orography, and secondary circulations such as mountain-valley breezes create daily cycles of circulation and dispersion regimes that, according to \textcite{castelli}, cannot be treated with simplified models or parameterisations. The pipeline therefore cannot reproduce wind-aligned plumes, transport along the valley, or the daily alternation of dispersion conditions. The diffusion coefficients add a further simplification: they are constant in space and time, and the values for the gases and for PM$_{2.5}$ differ only because they were tuned to different targets. Turbulent diffusion is a property of the turbulence rather than of the fluid and can vary greatly within a single flow~\citep{Tirabassi2009}. The horizontal coefficients are therefore best understood as calibration surrogates for processes the model does not fully resolve, rather than as physical parameters. For PM$_{2.5}$ a first-order loss term stands in for deposition and other removal (Section~\ref{sec:meth-dispersion}), as a single tuned rate rather than a physically resolved process. The vertical diffusivity $K_z(z)$ follows Monin--Obukhov similarity theory in the same spirit as \textcite{zhao2024}, but with fixed rather than hourly, site-derived $u_*$ and $L$; the theory itself assumes flat, horizontally homogeneous terrain, so adopting it is an improvement in structure rather than in physical validity for this alpine setting. Atmospheric chemistry and vehicle-induced turbulence, which \textcite{WOODWARD2019} show to matter at low wind speeds, are not represented. At the domain edges, the Dirichlet boundaries prevent pollutants emitted outside a village domain, for instance on neighbouring stretches of the A2 or in adjacent villages, from ever entering it, and the background itself is spatially uniform, constant in time, and identical on both dates, so seasonal differences in ambient concentrations between July and November are not represented.

\paragraph{Exposure.} The exposure stage produces an index, not a health outcome. The \ac{AQI} is the \ac{EPA}'s tool for communicating air quality, an index value of 100 generally corresponds to the level of the short-term US national ambient air quality standard, and its focus is short-term exposure~\citep{usepa_aqi_2026}. The summed sub-indices and the excess index are therefore relative measures for comparing villages, dates, and scenarios within this study, not statements of health risk. Concentrations are read at 0.3 to 1.2\,m, one layer above the ground into which vehicles emit (Appendix~\ref{app:tech-aqi}). Residents are assigned the outdoor concentration at their home around the clock. The 1-hour NO$_2$ sub-index is evaluated from hourly snapshots rather than hourly means, so short peaks between snapshots are missed while a snapshot that coincides with a peak is over-weighted. Finally, SynPop shuffles home locations among real addresses within clusters of at least 1500 persons~\citep{ARE2024SynPop}, so occupied addresses are preserved but not which resident lives at each of them.

\paragraph{Evaluation and computational cost.} No output of the pipeline has been compared against measurements in the corridor: the national monitoring network has no station there~\citep{nabel2025}, and the valley stations of the cantonal air quality office were used to set the calibration targets (Appendix~\ref{app:params}), so they cannot serve as independent validation. Where such data become available, point-to-point agreement should not be expected, since model output represents volume averages while observations are instantaneous single-point values~\citep{castelli}. The horizontal diffusion coefficients and the PM$_{2.5}$ loss rate were set by a scripted grid search against literature and station targets (Appendix~\ref{app:params}), so absolute concentrations carry no physical meaning; they are tuned to fall within a plausible range for the valley (between the measured background and near-road levels of Appendix~\ref{app:params}) and have not been validated. No formal sensitivity or uncertainty analysis has been carried out. Computational cost, in contrast, does not limit transferability: on a single laptop (AMD Ryzen 7 PRO 7840HS, 32\,GiB of memory), a full simulated day for the largest village, Schattdorf, took roughly two hours. Because emissions are queried through \ac{TraCI} for every active vehicle at every time-step, this cost grows with the number of vehicles in the domain, which is largest on precisely the congested days of greatest interest.

\subsection{Implications for the case-study results}
\label{sec:disc-implications}

Taken together, these limitations mean that the values reported in Section~\ref{sec:results} demonstrate what the pipeline can resolve rather than estimate exposure in the corridor, and they bear unevenly on the individual findings. The dominance of NO$_2$ in the traffic-attributable excess is the finding most exposed to modelling choices; with a primary NO$_2$/NO$_x$ ratio of 0.3 it gives way to PM$_{2.5}$ in most villages (Section~\ref{sec:res-villages}). It depends on the fleet, in which the diesel light-duty and truck types emit several times the NO$_x$ of the petrol cars, on the identification of NO$_x$ with NO$_2$, on the pollutant-specific calibration targets, whose sources differ between pollutants (Appendix~\ref{app:params}), on the absence of non-exhaust PM$_{2.5}$, and on the AQI breakpoints, which are lenient for CO. The ranking of villages by mean excess, in contrast, is unchanged under that scaling on both dates, but it depends on the heuristic local-traffic rules, most visibly in Schattdorf, where the weekday and weekend spawn profiles decide whether the busy or the quiet day is worse. The rerouting response was obtained in a single village whose rerouted vehicles cannot leave the domain. Absolute concentrations scale with heuristic diffusion coefficients. Each scenario was simulated once, Scenario~C on both dates, so neither the stochastic variability of local traffic nor day-to-day variability is quantified. We expect the qualitative structure of the results to be less sensitive to these choices than any individual value, but this has not been tested.

\subsection{Extending the pipeline}
\label{sec:disc-extensions}

Because the four stages communicate only through per-vehicle emissions, a concentration grid, and a population table, most of the limitations above can be addressed by replacing one stage without changing the others. The most consequential steps are hourly, site-derived $u_*$ and $L$ in place of the fixed values used here, and a wind field coupled to the diffusion coefficients rather than treated as an independent sensitivity check (Section~\ref{sec:windsens}); a fleet with explicit emission classes and a slope-aware emission model; NO--NO$_2$--O$_3$ chemistry in place of the identification of NO$_x$ with NO$_2$; and repeated stochastic realisations to quantify variability. Real terrain, multi-village domains, and activity-based exposure would follow. On the numerical side, higher-order time-stepping and more parallelisation would allow finer resolution.

\section{Conclusion}
\label{sec:conclusion}

We presented a coupled pipeline that links microscopic traffic simulation, per-vehicle emissions, three-dimensional implicit finite-difference diffusion, and resident-level exposure over a synthetic population, and released it as open-source software. Applied to six villages of the Gotthard corridor on a high- and a low-traffic day, it resolves the complete chain from moving vehicles to the air at residents' homes for the residential areas of six villages, with a full simulated day for the largest village taking roughly two hours on a single laptop.
 
The case study yields three methodological lessons. An \ac{AQI} in a low-background setting has to be decomposed into ambient background and traffic-attributable excess before settlements or days can be compared at all. Exposure at dwellings is not ordered by population: the traffic-attributable burden per resident is highest where diverted traffic passes close to dwellings, as in Wassen, while in Schattdorf, the largest village, it is set mainly by residents' own trips and hardly differs between the busy and the quiet day; this ordering is robust to the treatment of NO$_2$, whereas the dominant pollutant is not. And because traffic is simulated rather than prescribed, the pipeline is in principle able to represent traffic scenarios before they are implemented; as an illustration, in one run for Schattdorf, residential NO$_2$ fell with dynamic rerouting, with most of the change reached at the lowest share tested.
 
These findings are conditional on simplifications, chief among them the omission of wind in a valley whose dispersion is shaped by mountain-valley circulations, heuristic diffusion coefficients, a uniform vehicle fleet, and the absence of local evaluation data. The modular structure allows each of them to be relaxed independently. For the canton of Uri, which has already introduced ramp closures and speed reductions against diversion onto cantonal roads~\citep{KantonUri2026}, the pipeline could, in future work, be used to explore how such measures change the air that residents breathe, and not only how they change traffic; the measures themselves were not simulated here.

\section*{Data and code availability}
The full pipeline, together with the traffic patterns used in the case study (the through-traffic routes for both dates), is available at \url{https://github.com/cedriczeiter/zeiter_coupled_pipeline}. The Swiss Synthetic Population (SynPop) data contain person-level records and are not redistributed. They are used only in the exposure (AQI) step and in the parametrisation of the randomised local-traffic module; the traffic patterns released with the code are what is needed to rerun the traffic, emission and dispersion stages.

\section*{Acknowledgements}
The authors thank Manon Bertola for the corridor traffic simulation and for providing the traffic microsimulation models for two case-study days on request, as part of her master's thesis at ETH Zürich. 
We thank Jan Meier of the canton of Uri for guiding the field trip to the region, and Werner Gnos of the Uri cantonal police and Jonathan Imhof, press representative of the petition against detour traffic, for the interviews that informed our understanding of the situation on the ground, and Jan Lordieck for organising the contacts and meetings with these stakeholders. We thank the Swiss Federal Office for Spatial Development (ARE) and the Swiss Federal Railways (SBB), who commissioned the Swiss Synthetic Population 2022. Road network and building data are \copyright{} OpenStreetMap contributors. The terrain relief in Figure~\ref{fig:all_villages_squares} is derived from the Terrain Tiles dataset on the Registry of Open Data on AWS (Mapzen from \url{https://registry.opendata.aws/terrain-tiles}), and the outlines of Switzerland and its cantons and lakes are from Natural Earth (\url{https://www.naturalearthdata.com}). This research received no external funding.

%%%%%%%%%%%%% REFERENCES
\defbibheading{bibliography}[\bibname]{}
\renewcommand*{\bibfont}{\scriptsize\emergencystretch=3em}

\clearpage
\section*{References}
\printbibliography

@article{Abou-Senna2013,
	title        = {Using a traffic simulation model ({VISSIM}) with an emissions model ({MOVES}) to predict emissions from vehicles on a limited-access highway},
	author       = {Hatem Abou-Senna and Essam Radwan and Kurt Westerlund and C. David Cooper},
	year         = 2013,
	journal      = {Journal of the Air \& Waste Management Association},
	publisher    = {Taylor \& Francis},
	volume       = 63,
	number       = 7,
	pages        = {819--831},
	doi          = {10.1080/10962247.2013.795918}
}

@techreport{AfUUri2025,
	title        = {Immissionsmessungen Kanton Uri: {\"U}bersicht {\"u}ber die Immissionsmessungen im Kanton Uri mit Fokus auf das Jahr 2024},
	author       = {Wey, Sebastian and Engbersen, Nadine and Ruckstuhl, Christian},
	year         = 2025,
	month        = oct,
	address      = {Luzern},
	url          = {https://www.ur.ch/publikationen/7711},
	institution  = {inNET Monitoring AG, on behalf of Amt f{\"u}r Umwelt, Kanton Uri},
}

@dataset{ARE2024SynPop,
	title        = {Synthetische Bevölkerung der Schweiz 2022 ({SynPop} 2022)},
	author       = {{Schweizerische Bundesbahnen (SBB)} and {Bundesamt für Raumentwicklung (ARE)}},
	year         = 2024,
	publisher    = {FORS},
	doi          = {10.48573/mthb-qs95},
	version      = {1.0},
}

@online{astra_2025_gotthard_close,
	title        = {{A2P} Gotthardpass und Schöllenen: Wintersperre für Strassen- und Langsamverkehr},
	author       = {{Bundesamt für Strassen (ASTRA)}},
	url          = {https://www.astra.admin.ch/de/a2p-gotthardpass-und-schoellenen-wintersperre-fuer-strassen-und-langsamverkehr},
	urldate      = {2026-09-13},
	date         = {2025-11-05},
}

@online{astra_2026_gotthard_open,
	title        = {{A2P} Gotthard-Passstrasse - Öffnung des Passes nach der Wintersperre},
	author       = {{Bundesamt für Strassen (ASTRA)}},
	url          = {https://www.astra.admin.ch/de/a2p-gotthard-passstrasse-oeffnung-des-passes-nach-der-wintersperre},
	urldate      = {2026-09-13},
	date         = {2026-05-05},
}

@online{ASTRA202120Jahre,
	title        = {20 Jahre nach dem Brand},
	author       = {{Bundesamt für Strassen (ASTRA)}},
	year         = 2021,
	url          = {https://www.astra.admin.ch/de/20-jahre-nach-dem-brand},
	urldate      = {2026-09-13},
}

@techreport{ASTRA2025Faktenblatt,
	title        = {Faktenblatt Gotthard-Strassentunnel: Unf{\"a}lle und Verunfallte im Gotthard-Strassentunnel 2001--2024},
	author       = {{Bundesamt f{\"u}r Strassen (ASTRA)}},
	year         = 2025,
	month        = oct,
	number       = {ASTRA-D-313E3401/825},
	url          = {https://www.astra.admin.ch/dam/de/sd-web/0gsDw9u1oGeL/faktenblatt_gotthard-strassentunnel.pdf},
	urldate      = {2026-05-27},
	institution  = {Bundesamt f{\"u}r Strassen ASTRA}
}

@techreport{ASTRA2025Jahresbericht,
	title        = {Verkehrsentwicklung und Verkehrsfluss 2024},
	author       = {{Bundesamt f{\"u}r Strassen (ASTRA)}},
	year         = 2025,
	month        = jul,
	url          = {https://www.astra.admin.ch/dam/de/sd-web/Yh-UTVfv1ZWk/Verkehrsentwicklung_und_Verf%C3%BCgbarkeit_der_Nationalstrassen_Jahresbericht_2024.pdf},
	urldate      = {2026-05-27},
	institution  = {Bundesamt f{\"u}r Strassen ASTRA}
}

@techreport{BAFU2025Luftqualitaet,
	title        = {Luftqualität 2024: Messresultate des Nationalen Beobachtungsnetzes für Luftfremdstoffe ({NABEL})},
	author       = {{Bundesamt für Umwelt (BAFU)} and {Eidgenössische Materialprüfungs- und Forschungsanstalt (Empa)}},
	year         = 2025,
	address      = {Bern},
	series       = {Umwelt-Zustand},
	url          = {https://www.bafu.admin.ch/de/publication?id=85q0ngZQNJf1},
	institution  = {Bundesamt für Umwelt (BAFU)},
}

@report{Barth2000,
	title        = {The Development of a Comprehensive Modal Emissions Model},
	author       = {Barth, Matthew and An, Feng and Younglove, Theodore and Scora, George and Levine, Carrie and Ross, Marc and Wenzel, Thomas},
	location     = {Washington, DC},
	type         = {NCHRP Web-Only Document, Final Report Project 25-11},
	institution  = {National Cooperative Highway Research Program},
	date         = 2000
}

@inproceedings{behrisch2024,
	title        = {Comparing and Parameterizing the Electrical Energy Consumption Models in {SUMO}},
	author       = {Behrisch, Michael and Fl{\"o}tter{\"o}d, Yun-Pang and Wagner, Peter},
	year         = 2024,
	booktitle    ={{SUMO} Conference Proceedings},
	volume       = 5,
	doi          = {10.52825/scp.v5i.1012}
}

@article{Berkowicz2000,
	title        = {{OSPM} -- A Parameterised Street Pollution Model},
	author       = {Berkowicz, Ruwim},
	year         = 2000,
	month        = nov,
	day          = {01},
	journal      = {Environmental Monitoring and Assessment},
	volume       = 65,
	pages        = {323--331},
	doi          = {10.1023/A:1006448321977},
	issn         = {1573-2959}
}

@article{Boogaard2022,
	title        = {Long-term exposure to traffic-related air pollution and selected health outcomes: A systematic review and meta-analysis},
	author       = {H. Boogaard and A.P. Patton and R.W. Atkinson and J.R. Brook and H.H. Chang and D.L. Crouse and J.C. Fussell and G. Hoek and B. Hoffmann and R. Kappeler and M. {Kutlar Joss} and M. Ondras and S.K. Sagiv and E. Samoli and R. Shaikh and A. Smargiassi and A.A. Szpiro and E.D.S. {Van Vliet} and D. Vienneau and J. Weuve and F.W. Lurmann and F. Forastiere},
	year         = 2022,
	journal      = {Environment International},
	volume       = 164,
	pages        = 107262,
	doi          = {10.1016/j.envint.2022.107262},
	issn         = {0160-4120}
}

@article{businger1971,
	title        = {Flux-Profile Relationships in the Atmospheric Surface Layer},
	author       = {J. A.  Businger and J. C.  Wyngaard and Y.  Izumi and E. F.  Bradley},
	year         = 1971,
	journal      = {Journal of the Atmospheric Sciences},
	publisher    = {American Meteorological Society},
	address      = {Boston MA, USA},
	volume       = 28,
	number       = 2,
	pages        = {181--189},
	doi          = {10.1175/1520-0469(1971)028<0181:FPRITA>2.0.CO;2}
}

@online{Campax2022,
	title        = {Urner D{\"o}rfer vor Ausweichverkehr sch{\"u}tzen!},
	author       = {{IG Uri}},
	year         = 2022,
	url          = {https://act.campax.org/petitions/stopp-ausweichverkehr-a2},
	urldate      = {2026-05-27},
    
}

@article{carslaw2005,
	title        = {Estimations of road vehicle primary {NO\textsubscript{2}} exhaust emission fractions using monitoring data in London},
	author       = {David C. Carslaw and Sean D. Beevers},
	year         = 2005,
	journal      = {Atmospheric Environment},
	volume       = 39,
	number       = 1,
	pages        = {167--177},
	doi          = {10.1016/j.atmosenv.2004.08.053},
	issn         = {1352-2310}
}

@article{castelli,
	title        = {Modelling the meteorology and traffic pollutant dispersion in highly complex terrain: The {ALPNAP} Alpine Space Project},
	author       = {Trini Castelli, Silvia and Belfiore, Giovanni and Anfossi, D. and Elampe, Elisa and Clemente, Monica},
	year         = 2011,
    pages = {235--243},
	month        = {02},
	journal      = {Int. J. of Environment and Pollution},
	volume       = 44,
	doi          = {10.1504/IJEP.2011.038423}
}

@article{Chen2017,
	title        = {Living near major roads and the incidence of dementia, Parkinson's disease, and multiple sclerosis: a population-based cohort study},
	author       = {Chen, Hong and Kwong, Jeffrey C. and Copes, Ray and Tu, Karen and Villeneuve, Paul J. and van Donkelaar, Aaron and Hystad, Perry and Martin, Randall V. and Murray, Brian J. and Jessiman, Barry and Wilton, Andrew S. and Kopp, Alexander and Burnett, Richard T.},
	year         = 2017,
	month        = feb,
	day          = 18,
	journal      = {The Lancet},
	publisher    = {Elsevier},
	volume       = 389,
	number       = 10070,
	pages        = {718--726},
	doi          = {10.1016/S0140-6736(16)32399-6},
}

@article{Cimorelli2005,
	title        = {{AERMOD}: A Dispersion Model for Industrial Source Applications. {Part I}: General Model Formulation and Boundary Layer Characterization},
	author       = {Alan J. Cimorelli and Steven G. Perry and Akula Venkatram and Jeffrey C. Weil and Robert J. Paine and Robert B. Wilson and Russell F. Lee and Warren D. Peters and Roger W. Brode},
	year         = 2005,
	journal      = {Journal of Applied Meteorology},
	publisher    = {American Meteorological Society},
	address      = {Boston MA, USA},
	volume       = 44,
	number       = 5,
	pages        = {682--693},
	doi          = {10.1175/JAM2227.1}
}

@article{Clements2009,
	title        = {Air pollutant concentrations near three Texas roadways, part II: Chemical characterization and transformation of pollutants},
	author       = {Andrea L. Clements and Yuling Jia and Allison Denbleyker and Elena McDonald-Buller and Matthew P. Fraser and David T. Allen and Donald R. Collins and Edward Michel and Jayanth Pudota and David Sullivan and Yifang Zhu},
	year         = 2009,
	month        = sep,
	journal      = {Atmospheric Environment},
	volume       = 43,
	number       = 30,
	pages        = {4523--4534},
	doi          = {10.1016/j.atmosenv.2009.06.044},
	issn         = {1352-2310}
}

@article{colberg2005,
	title        = {Comparison of a road traffic emission model ({HBEFA}) with emissions derived from measurements in the Gubrist road tunnel, Switzerland},
	author       = {Christina A. Colberg and Bruno Tona and Werner A. Stahel and Markus Meier and Johannes Staehelin},
	year         = 2005,
	journal      = {Atmospheric Environment},
	volume       = 39,
	number       = 26,
	pages        = {4703--4714},
	doi          = {10.1016/j.atmosenv.2005.04.020},
	issn         = {1352-2310}
}

@article{Demiraslan2025,
	title        = {Effects of topographic variables on traffic-related pollutant concentrations: comparison of {AERMOD} and {CAL3QHCR} models},
	author       = {Demirarslan, Kazım O.},
	year         = 2025,
	journal      = {Frontiers in Environmental Science},
	volume       = {13},
	doi          = {10.3389/fenvs.2025.1577330},
}

@article{denazelle2013,
	title        = {Improving estimates of air pollution exposure through ubiquitous sensing technologies},
	author       = {Audrey {de Nazelle} and Edmund Seto and David Donaire-Gonzalez and Michelle Mendez and Jaume Matamala and Mark J. Nieuwenhuijsen and Michael Jerrett},
	year         = 2013,
	journal      = {Environmental Pollution},
	volume       = 176,
	pages        = {92--99},
	doi          = {10.1016/j.envpol.2012.12.032},
	issn         = {0269-7491}
}

@article{DIAS2016253,
	title        = {Integrated modelling approach for the evaluation of low emission zones},
	author       = {Daniela Dias and Oxana Tchepel and António Pais Antunes},
	year         = 2016,
	journal      = {Journal of Environmental Management},
	volume       = 177,
	pages        = {253--263},
	doi          = {10.1016/j.jenvman.2016.04.031},
	issn         = {0301-4797}
}

@article{DIB2025104791,
	title        = {Deep learning approach to predict microscopic pollutant emissions from mesoscopic traffic simulations},
	author       = {Abdelkader Dib and Milos Balac and Antonio Sciarretta},
	year         = 2025,
	journal      = {Transportation Research Part D: Transport and Environment},
	volume       = 146,
	pages        = 104791,
	doi          = {10.1016/j.trd.2025.104791},
	issn         = {1361-9209}
}

@article{Duyzer2015,
	title        = {Representativeness of air quality monitoring networks},
	author       = {Jan Duyzer and Dick {van den Hout} and Peter Zandveld and Sjoerd {van Ratingen}},
	year         = 2015,
	journal      = {Atmospheric Environment},
	volume       = 104,
	pages        = {88--101},
	doi          = {10.1016/j.atmosenv.2014.12.067},
	issn         = {1352-2310}
}

@techreport{Eckhoff1995,
	title        = {Addendum to the User's Guide to {CAL3QHC} Version 2.0 ({CAL3QHCR} User's Guide)},
	author       = {Eckhoff, Peter A. and Braverman, Thomas N.},
	year         = 1995,
	month        = sep,
	address      = {Research Triangle Park, NC},
	url          = {https://www.epa.gov/scram/air-quality-dispersion-modeling-alternative-models},
	institution  = {U.S. Environmental Protection Agency}
}

@techreport{emep2024,
	title        = {Passenger cars, light commercial trucks, heavy-duty vehicles including buses and motorcycles},
	author       = {Ntziachristos, Leonidas and Samaras, Zissis},
	year         = 2024,
	number       = {NFR 1.A.3.b.i-iv},
	url          = {https://www.eea.europa.eu/publications/emep-eea-guidebook-2023},
	institution  = {European Environment Agency},
}

@inbook{Fellendorf2010,
	title        = {Microscopic Traffic Flow Simulator {VISSIM}},
	author       = {Fellendorf, Martin and Vortisch, Peter},
	year         = 2010,
	booktitle    = {Fundamentals of Traffic Simulation},
	publisher    = {Springer New York},
	address      = {New York, NY},
	pages        = {63--93},
	doi          = {10.1007/978-1-4419-6142-6_2},
	isbn         = {978-1-4419-6142-6},
	editor       = {Barcel{\'o}, Jaume}
}

@article{Fuller2022,
	title        = {Pollution and health: a progress update},
	author       = {Fuller, Richard and Landrigan, Philip J. and Balakrishnan, Kalpana and Bathan, Glynda and Bose-O'Reilly, Stephan and Brauer, Michael and Caravanos, Jack and Chiles, Tom and Cohen, Aaron and Corra, Lilian and Cropper, Maureen and Ferraro, Greg and Hanna, Jill and Hanrahan, David and Hu, Howard and Hunter, David and Janata, Gloria and Kupka, Rachael and Lanphear, Bruce and Lichtveld, Maureen and Martin, Keith and Mustapha, Adetoun and Sanchez-Triana, Ernesto and Sandilya, Karti and Schaefli, Laura and Shaw, Joseph and Seddon, Jessica and Suk, William and T{\'e}llez-Rojo, Martha Mar{\'i}a and Yan, Chonghuai},
	year         = 2022,
	month        = jun,
	day          = {01},
	journal      = {The Lancet Planetary Health},
	publisher    = {Elsevier},
	volume       = 6,
	pages        = {e535--e547},
	doi          = {10.1016/S2542-5196(22)00090-0},
	issn         = {2542-5196}
}

@article{haklay2008osm,
	title        = {{OpenStreetMap}: User-Generated Street Maps},
	author       = {Haklay, Mordechai and Weber, Patrick},
	year         = 2008,
	journal      = {IEEE Pervasive Computing},
	volume       = 7,
	number       = 4,
	pages        = {12--18},
	doi          = {10.1109/MPRV.2008.80}
}

@article{HATZOPOULOU2010315,
	title        = {Linking an activity-based travel demand model with traffic emission and dispersion models: Transport’s contribution to air pollution in Toronto},
	author       = {M. Hatzopoulou and E. J. Miller},
	year         = 2010,
	journal      = {Transportation Research Part D: Transport and Environment},
	volume       = 15,
	number       = 6,
	pages        = {315--325},
	doi          = {10.1016/j.trd.2010.03.007},
	issn         = {1361-9209}
}

@report{Hausberger2009,
	title        = {Emission Factors from the Model {PHEM} for the {HBEFA} Version 3},
	author       = {Hausberger, Stefan and Rexeis, Martin and Zallinger, Michael and Luz, Raphael},
	location     = {Graz},
	institution  = {Graz University of Technology, Institute for Internal Combustion Engines and Thermodynamics},
	date         = 2009
}

@online{hbefa_web,
	title        = {{HBEFA}: Handbook Emission Factors for Road Transport},
	author       = {{INFRAS}},
	year         = 2025,
	url          = {https://www.hbefa.net/},
	urldate      = {2026-09-23}
}

@InProceedings{hirschmann2010,
author={Hirschmann, K.
and Zallinger, M.
and Fellendorf, M.
and Hausberger, S.},
title={A new method to calculate emissions with simulated traffic conditions},
  booktitle={13th International {IEEE} Conference on Intelligent Transportation Systems}, 
year={2010},
pages={33--38},
issn={2153-0017},
doi={10.1109/ITSC.2010.5625030},
}

@article{HOFER20181,
	title        = {Large scale simulation of {CO\textsubscript{2}} emissions caused by urban car traffic: An agent-based network approach},
	author       = {Christian Hofer and Georg Jäger and Manfred Füllsack},
	year         = 2018,
	journal      = {Journal of Cleaner Production},
	volume       = 183,
	pages        = {1--10},
	doi          = {10.1016/j.jclepro.2018.02.113},
	issn         = {0959-6526}
}

@book{Horni2016,
	title        = {The Multi-Agent Transport Simulation {MATSim}},
	year         = 2016,
	publisher    = {Ubiquity Press},
	doi          = {10.5334/baw},
	isbn         = 9781909188754,
	editor       = {Horni, Andreas and Nagel, Kai and Axhausen, Kay W.}
}

@article{Hu2023,
	title        = {Changing trends in the air pollution--related disease burden from 1990 to 2019 and its predicted level in 25 years},
	author       = {Hu, Wan and Fang, Lanlan and Zhang, Hengchuan and Ni, Ruyu and Pan, Guixia},
	year         = 2023,
	month        = jan,
	day          = {01},
	journal      = {Environmental Science and Pollution Research},
	volume       = 30,
	pages        = {1761--1773},
	doi          = {10.1007/s11356-022-22318-z},
	issn         = {1614-7499}
}

@article{HUANG2009,
	title        = {A comparative study of the emissions by road maintenance works and the disrupted traffic using life cycle assessment and micro-simulation},
	author       = {Yue Huang and Roger Bird and Margaret Bell},
	year         = 2009,
	journal      = {Transportation Research Part D: Transport and Environment},
	volume       = 14,
	number       = 3,
	pages        = {197--204},
	doi          = {10.1016/j.trd.2008.12.003},
	issn         = {1361-9209}
}

@article{HULSMANN2014732,
	title        = {Modelling traffic and air pollution in an integrated approach – the case of Munich},
	author       = {Friederike Hülsmann and Regine Gerike and Matthias Ketzel},
	year         = 2014,
	journal      = {Urban Climate},
	volume       = 10,
	pages        = {732--744},
	doi          = {10.1016/j.uclim.2014.01.001},
	issn         = {2212-0955}
}

@article{Jang2025,
	title        = {Tools for estimating traffic emissions considering spatial scales and mitigation strategies: A systematic review},
	author       = {Saewoom Jang and Junseo Park and Gahyeong Yu and Jinuk Hwang},
	year         = 2025,
	journal      = {International Journal of Sustainable Transportation},
	publisher    = {Taylor \& Francis},
	volume       = 19,
	number       = 5,
	pages        = {431--445},
	doi          = {10.1080/15568318.2025.2502558}
}

@online{KantonUri2026,
	title        = {Staumanagement Uri: Massnahmen bis zur {\"O}ffnung des Gotthardpasses 2026},
	author       = {{Kanton Uri}},
	year         = 2026,
	url          = {https://www.ur.ch/mmdirektionen/133927},
	urldate      = {2026-05-27},

}

@article{Lelieveld2015,
	title        = {The contribution of outdoor air pollution sources to premature mortality on a global scale},
	author       = {Lelieveld, J. and Evans, J. S. and Fnais, M. and Giannadaki, D. and Pozzer, A.},
	year         = 2015,
	month        = sep,
	day          = {01},
	journal      = {Nature},
	volume       = 525,
	number       = 7569,
	pages        = {367--371},
	doi          = {10.1038/nature15371},
	issn         = {1476-4687}
}

@article{Lelieveld2023,
	title        = {Air pollution deaths attributable to fossil fuels: observational and modelling study},
	author       = {Lelieveld, Jos and Haines, Andy and Burnett, Richard and Tonne, Cathryn and Klingmüller, Klaus and Münzel, Thomas and Pozzer, Andrea},
	year         = 2023,
	journal      = {{BMJ}},
	volume       = 383,
	pages        = {e077784},
	doi          = {10.1136/bmj-2023-077784},
	issn         = {0959-8138}
}

@article{LIU20111154,
	title        = {Traffic and emission simulation in China based on statistical methodology},
	author       = {Huan Liu and Kebin He and Matthew Barth},
	year         = 2011,
	journal      = {Atmospheric Environment},
	volume       = 45,
	number       = 5,
	pages        = {1154--1161},
	doi          = {10.1016/j.atmosenv.2010.10.051},
	issn         = {1352-2310}
}

@inproceedings{Lopez2018,
  author={Alvarez Lopez, Pablo and Behrisch, Michael and Bieker-Walz, Laura and Erdmann, Jakob and Flötteröd, Yun-Pang and Hilbrich, Robert and Lücken, Leonhard and Rummel, Johannes and Wagner, Peter and Wiessner, Evamarie},
  booktitle={2018 21st International Conference on Intelligent Transportation Systems ({ITSC})}, 
  title={Microscopic Traffic Simulation using {SUMO}}, 
  year={2018},
  pages={2575--2582},
  doi={10.1109/ITSC.2018.8569938}
  }

@article{Ma2012,
	title        = {An Evaluation of Microscopic Emission Models for Traffic Pollution Simulation Using On-board Measurement},
	author       = {Ma, Xiaoliang and Lei, Wei and Andr{\'e}asson, Ingmar and Chen, Hui},
	year         = 2012,
	month        = aug,
	day          = {01},
	journal      = {Environmental Modeling {\&} Assessment},
	volume       = 17,
	pages        = {375--387},
	doi          = {10.1007/s10666-011-9296-9},
}

@article{madziel2023,
	title        = {Vehicle Emission Models and Traffic Simulators: A Review},
	author       = {Mądziel, Maksymilian},
	year         = 2023,
	journal      = {Energies},
	volume       = 16,
	number       = 9,
	doi          = {10.3390/en16093941},
	issn         = {1996-1073},
	article-number = 3941
}

@article{monn1998,
	title        = {Personal exposure to nitrogen dioxide in Switzerland},
	author       = {C. Monn and O. Brändli and C. Schindler and U. Ackermann-Liebrich and P. Leuenberger},
	year         = 1998,
	journal      = {Science of The Total Environment},
	volume       = 215,
	number       = 3,
	pages        = {243--251},
	doi          = {10.1016/S0048-9697(98)00124-7},
	issn         = {0048-9697}
}

@article{Mukherjee2020,
	title        = {Influence of roadway emissions on near-road {PM\textsubscript{2.5}}: Monitoring data analysis and implications},
	author       = {Mukherjee, Anondo and McCarthy, Michael C. and Brown, Steven G. and Huang, ShihMing and Landsberg, Karin and Eisinger, Douglas S.},
	year         = 2020,
	month        = sep,
	journal      = {Transportation Research Part D: Transport and Environment},
	volume       = 86,
	pages        = 102442,
	doi          = {10.1016/j.trd.2020.102442},
	issn         = {1361-9209}
}

@online{nabel2025,
	title        = {National Air Pollution Monitoring Network ({NABEL})},
	author       = {{Federal Office for the Environment (FOEN)}},
	year         = 2025,
	month        = dec,
	day          = 5,
	url          = {https://www.bafu.admin.ch/en/national-air-pollution-monitoring-network-nabel},
	urldate      = {2026-09-01}
}

@inproceedings{Ntziachristos2009,
	title        = {{COPERT}: A European Road Transport Emission Inventory Model},
	author       = {Ntziachristos, Leonidas and Gkatzoflias, Dimitrios and Kouridis, Chariton and Samaras, Zissis},
	year         = 2009,
	booktitle    = {Information Technologies in Environmental Engineering},
	publisher    = {Springer},
	address      = {Berlin, Heidelberg},
	pages        = {491--504},
	doi          = {10.1007/978-3-540-88351-7_37},
	isbn         = {978-3-540-88351-7},
	editor       = {Athanasiadis, Ioannis N. and Rizzoli, Andrea E. and Mitkas, Pericles A. and G{\'o}mez, Jorge Marx}
}

@article{odriscoll2016,
	title        = {A Portable Emissions Measurement System ({PEMS}) study of {NO\textsubscript{x}} and primary {NO\textsubscript{2}} emissions from Euro 6 diesel passenger cars and comparison with {COPERT} emission factors},
	author       = {Rosalind O'Driscoll and Helen M. ApSimon and Tim Oxley and Nick Molden and Marc E. J. Stettler and Aravinth Thiyagarajah},
	year         = 2016,
	journal      = {Atmospheric Environment},
	volume       = 145,
	pages        = {81--91},
	doi          = {10.1016/j.atmosenv.2016.09.021},
	issn         = {1352-2310}
}

@article{pant2013,
	title        = {Estimation of the contribution of road traffic emissions to particulate matter concentrations from field measurements: A review},
	author       = {Pallavi Pant and Roy M. Harrison},
	year         = 2013,
	journal      = {Atmospheric Environment},
	volume       = 77,
	pages        = {78--97},
	doi          = {10.1016/j.atmosenv.2013.04.028},
	issn         = {1352-2310}
}

@article{Pantusheva2022,
	title        = {Air Pollution Dispersion Modelling in Urban Environment Using {CFD}: A Systematic Review},
	author       = {Pantusheva, Mariya and Mitkov, Radostin and Hristov, Petar O. and Petrova-Antonova, Dessislava},
	year         = 2022,
	month        = oct,
	journal      = {Atmosphere},
	publisher    = {MDPI},
	volume       = 13,
	number       = 10,
	pages        = 1640,
	doi          = {10.3390/atmos13101640},
	issn         = {2073-4433}
}

@article{PICCOLI2024,
	title        = {A bottom-up modelling chain to evaluate the impact of urban road transport policies on air quality and human health},
	author       = {Andrea Piccoli and Valentina Agresti and Marco Bedogni and Giovanni Lonati and Guido Pirovano},
	year         = 2024,
	journal      = {Urban Climate},
	volume       = 55,
	pages        = 101893,
	doi          = {10.1016/j.uclim.2024.101893},
	issn         = {2212-0955}
}

@online{ptv_visum,
	title        = {{PTV Visum}: Transport Planning Software},
	author       = {{PTV Group}},
	url          = {https://www.ptvgroup.com/en/products/ptv-visum},
	urldate      = {2026-09-27}
}

@article{QUAASSDORFF2016,
	title        = {Microscale traffic simulation and emission estimation in a heavily trafficked roundabout in Madrid (Spain)},
	author       = {Christina Quaassdorff and Rafael Borge and Javier Pérez and Julio Lumbreras and David {de la Paz} and Juan Manuel {de Andrés}},
	year         = 2016,
	journal      = {Science of The Total Environment},
	volume       = {566-567},
	pages        = {416--427},
	doi          = {10.1016/j.scitotenv.2016.05.051},
	issn         = {0048-9697}
}

@article{Rakha2004,
	title        = {Development of {VT-Micro} model for estimating hot stabilized light duty vehicle and truck emissions},
	author       = {Hesham Rakha and Kyoungho Ahn and Antonio Trani},
	year         = 2004,
	journal      = {Transportation Research Part D: Transport and Environment},
	volume       = 9,
	number       = 1,
	pages        = {49--74},
	doi          = {10.1016/S1361-9209(03)00054-3},
	issn         = {1361-9209}
}

@manual{ramboll2026camx,
	title        = {User's Guide: Comprehensive Air Quality Model with Extensions},
	author       = {Yarwood, Greg and Wilson, Gary and Emery, Chris and Morris, Ralph},
	year         = 2026,
	month        = jul,
	url          = {https://www.camx.com/Files/CAMxUsersGuide_v7.40.pdf},
	organization = {Ramboll Americas Engineering Solutions, Inc.},
	edition      = {Version 7.40}
}

@article{RichmondBryant2017,
	title        = {Estimation of on-road {NO\textsubscript{2}} concentrations, {NO\textsubscript{2}}/{NO\textsubscript{x}} ratios, and related roadway gradients from near-road monitoring data},
	author       = {Richmond-Bryant, Jennifer and Owen, Chris Robert and Graham, Stephen and Snyder, Michelle and McDow, Stephen and Oakes, Michelle and Kimbrough, Sue},
	year         = 2017,
	month        = jun,
	journal      = {Air Quality, Atmosphere {\&} Health},
	volume       = 10,
	pages        = {611--625},
	doi          = {10.1007/s11869-016-0455-7},
}

@article{samake2022,
	title        = {A Mathematical and Numerical Framework for Traffic-Induced Air Pollution Simulation in Bamako},
	author       = {Samaké, Abdoulaye and Mahamane, Amadou and Alassane, Mahamadou and Diallo, Ouaténi},
	year         = 2022,
	journal      = {Computation},
	volume       = 10,
	number       = 5,
	doi          = {10.3390/computation10050076},
	issn         = {2079-3197},
	article-number = 76
}

@techreport{scire2000calpuff,
	title        = {A User's Guide for the {CALPUFF} Dispersion Model (Version 5)},
	author       = {Scire, Joseph S. and Strimaitis, David G. and Yamartino, Robert J.},
	year         = 2000,
	month        = jan,
	address      = {Concord, MA},
	institution  = {Earth Tech, Inc.}
}

@article{segura2025,
	title        = {Modelling of Air Pollution Dispersion Coupled with Traffic Flow Simulation, as a Digital Twin, for the Correct Implementation of Low Emission Zones: An Innovative Approach for Smart Cities},
	author       = {Antonio Segura Morilla and Samuel Pineiro Pena and David Cartelle Fernandez},
	year         = 2025,
	month        = dec,
	journal      = {PriMera Scientific Engineering},
	volume       = 7,
	number       = 6,
	pages        = {36--44},
	issn         = {2834-2550},
	url          = {https://primerascientific.com/pdf/psen/PSEN-07-239.pdf}
}

@article{SILVEIRA2023119481,
	title        = {A multiscale air quality and health risk modelling system: Design and application over a local traffic management case study},
	author       = {Carlos Silveira and Joana Ferreira and Ana I. Miranda},
	year         = 2023,
	journal      = {Atmospheric Environment},
	volume       = 294,
	pages        = 119481,
	doi          = {10.1016/j.atmosenv.2022.119481},
	issn         = {1352-2310}
}

@article{Smit2007,
	title        = {A new modelling approach for road traffic emissions: {VERSIT+}},
	author       = {Robin Smit and Richard Smokers and Elke Rabé},
	year         = 2007,
	journal      = {Transportation Research Part D: Transport and Environment},
	volume       = 12,
	number       = 6,
	pages        = {414--422},
	doi          = {10.1016/j.trd.2007.05.001},
	issn         = {1361-9209}
}

@article{smit2010,
	title        = {Validation of road vehicle and traffic emission models – A review and meta-analysis},
	author       = {Robin Smit and Leonidas Ntziachristos and Paul Boulter},
	year         = 2010,
	journal      = {Atmospheric Environment},
	volume       = 44,
	number       = 25,
	pages        = {2943--2953},
	doi          = {10.1016/j.atmosenv.2010.05.022},
	issn         = {1352-2310}
}

@article{So21042018,
	title        = {Estimating emissions based on the integration of microscopic traffic simulation and vehicle dynamics model},
	author       = {Jaehyun So and Nassim Motamedidehkordi and Yingjie Wu and Fritz Busch and Keechoo Choi},
	year         = 2018,
	journal      = {International Journal of Sustainable Transportation},
	publisher    = {Taylor \& Francis},
	volume       = 12,
	number       = 4,
	pages        = {286--298},
	doi          = {10.1080/15568318.2017.1363328}
}

@article{SONNENSCHEIN2025,
	title        = {Hybrid cellular automata-based air pollution model for traffic scenario microsimulations},
	author       = {Tabea S. Sonnenschein and Zhendong Yuan and Jibran Khan and Jules Kerckhoffs and Roel C. H. Vermeulen and Simon Scheider},
	year         = 2025,
	journal      = {Environmental Modelling \& Software},
	volume       = 186,
	pages        = 106356,
	doi          = {10.1016/j.envsoft.2025.106356},
	issn         = {1364-8152}
}

@article{Steinbacher2007,
	title        = {Perennial observations of molecular hydrogen ({H\textsubscript{2}}) at a suburban site in Switzerland},
	author       = {Steinbacher, Martin and Fischer, Andrea and Vollmer, Martin K. and Buchmann, Brigitte and Reimann, Stefan and Hueglin, Christoph},
	year         = 2007,
	month        = mar,
	journal      = {Atmospheric Environment},
	volume       = 41,
	number       = 10,
	pages        = {2111--2124},
	doi          = {10.1016/j.atmosenv.2006.10.075},
	issn         = {1352-2310}
}

@online{sumo_hbefa4,
	title        = {{HBEFA4}-based},
	author       = {{SUMO}},
	year         = 2024,
	url          = {https://sumo.dlr.de/docs/Models/Emissions/HBEFA4-based.html},
	urldate      = {2026-09-26}
}

@inbook{Tirabassi2009,
	title        = {Mathematical Air Pollution Models: Eulerian Models},
	author       = {Tirabassi, Tiziano},
	year         = 2009,
	booktitle    = {Air Pollution and Turbulence},
	publisher    = {CRC Press},
	pages        = {131--155},
	doi          = {10.1201/9781439847121-5},
	isbn         = 9780429110436,
	edition      = {First}
}

@article{Tu2018,
	title        = {Development of a hybrid modelling approach for the generation of an urban on-road transportation emission inventory},
	author       = {Ran Tu and Islam Kamel and An Wang and Baher Abdulhai and Marianne Hatzopoulou},
	year         = 2018,
	journal      = {Transportation Research Part D: Transport and Environment},
	volume       = 62,
	pages        = {604--618},
	doi          = {10.1016/j.trd.2018.04.011},
	issn         = {1361-9209}
}

@online{UrnerZeitung2023,
	title        = {Petition gegen Ausweichverkehr durch {Urner} {D{\"o}rfer} eingereicht},
	author       = {{Urner Zeitung}},
	year         = 2023,
	url          = {https://www.urnerzeitung.ch/zentralschweiz/uri/altdorf-petition-gegen-den-ausweichverkehr-3034-unterschriften-eingereicht-ld.2409937},
	urldate      = {2026-05-27},

}

@techreport{usepa_aqi_2026,
	title        = {Technical Assistance Document for the Reporting of Daily Air Quality – the Air Quality Index ({AQI})},
	author       = {{U.S. Environmental Protection Agency}},
	year         = 2026,
	month        = may,
	url          = {https://www.airnow.gov/publications/air-quality-index/technical-assistance-document-for-reporting-the-daily-aqi/},
}

@techreport{usepa2003mobile6,
	title        = {User's Guide to {MOBILE6.1} and {MOBILE6.2}: Mobile Source Emission Factor Model},
	author       = {{U.S. Environmental Protection Agency}},
	location     = {Ann Arbor, MI},
	number       = {EPA420-R-02-028},
	url          = {https://permanent.fdlp.gov/gpo134489/P1001DSD.pdf},
	type         = {Technical Report},
	date         = {2002-10}
}

@article{VanVliet1997,
	title        = {Motor Vehicle Exhaust and Chronic Respiratory Symptoms in Children Living near Freeways},
	author       = {Patricia {van Vliet} and Mirjam Knape and Jeroen {de Hartog} and Nicole Janssen and Hendrik Harssema and Bert Brunekreef},
	year         = 1997,
	journal      = {Environmental Research},
	volume       = 74,
	number       = 2,
	pages        = {122--132},
	doi          = {10.1006/enrs.1997.3757},
	issn         = {0013-9351}
}

@article{wang2023,
	title        = {Aggravated air pollution and health burden due to traffic congestion in urban China},
	author       = {Wang, P. and Zhang, R. and Sun, S. and Gao, M. and Zheng, B. and Zhang, D. and Zhang, Y. and Carmichael, G. R. and Zhang, H.},
	year         = 2023,
	journal      = {Atmospheric Chemistry and Physics},
	volume       = 23,
	number       = 5,
	pages        = {2983--2996},
	doi          = {10.5194/acp-23-2983-2023}
}

@inproceedings{wegener2008traci,
	title        = {{TraCI}: an interface for coupling road traffic and network simulators},
	author       = {Wegener, Axel and Pi{\'o}rkowski, Micha{\l} and Raya, Maxim and Hellbr{\"u}ck, Horst and Fischer, Stefan and Hubaux, Jean-Pierre},
	year         = 2008,
	booktitle    = {Proceedings of the 11th Communications and Networking Simulation Symposium},
	publisher    = {Association for Computing Machinery},
	series       = {CNS '08},
	pages        = {155--163},
	doi          = {10.1145/1400713.1400740},
	isbn         = 1565553187,
	numpages     = 9
}

@article{WOODWARD2019,
	title        = {A large eddy simulation of the dispersion of traffic emissions by moving vehicles at an intersection},
	author       = {Huw Woodward and Marc Stettler and Dimitrios Pavlidis and Elsa Aristodemou and Helen ApSimon and Christopher Pain},
	year         = 2019,
	journal      = {Atmospheric Environment},
	volume       = 215,
	pages        = 116891,
	doi          = {10.1016/j.atmosenv.2019.116891},
	issn         = {1352-2310}
}

@article{Yuchi2020,
	title        = {Road proximity, air pollution, noise, green space and neurologic disease incidence: a population-based cohort study},
	author       = {Yuchi, Weiran and Sbihi, Hind and Davies, Hugh and Tamburic, Lillian and Brauer, Michael},
	year         = 2020,
	month        = jan,
	day          = 21,
	journal      = {Environmental Health},
	volume       = 19,
	pages        = 8,
	doi          = {10.1186/s12940-020-0565-4},
}

@article{ZHANG2013,
	title        = {Air pollution and health risks due to vehicle traffic},
	author       = {Kai Zhang and Stuart Batterman},
	year         = 2013,
	journal      = {Science of The Total Environment},
	volume       = {450-451},
	pages        = {307--316},
	doi          = {10.1016/j.scitotenv.2013.01.074},
	issn         = {0048-9697}
}

@inproceedings{zhao2024,
	title        = {Integrated Simulation Platform for Quantifying the Traffic-Induced Environmental and Health Impacts},
	author       = {Zhao, Xuanpeng and Wu, Guoyuan and Venkatram, Akula and Luo, Ji and Hao, Peng and Boriboonsomsin, Kanok and Hu, Shaohua},
	year         = 2024,
	booktitle    = {2024 Forum for Innovative Sustainable Transportation Systems ({FISTS})},
	pages        = {1--8},
	doi          = {10.1109/FISTS60717.2024.10485596}
}

@article{ZHOU2015,
	title        = {Integrating a simplified emission estimation model and mesoscopic dynamic traffic simulator to efficiently evaluate emission impacts of traffic management strategies},
	author       = {Xuesong Zhou and Shams Tanvir and Hao Lei and Jeffrey Taylor and Bin Liu and Nagui M. Rouphail and H. {Christopher Frey}},
	year         = 2015,
	journal      = {Transportation Research Part D: Transport and Environment},
	volume       = 37,
	pages        = {123--136},
	doi          = {10.1016/j.trd.2015.04.013},
	issn         = {1361-9209}
}

@article{Zhu2002,
	title        = {Study of ultrafine particles near a major highway with heavy-duty diesel traffic},
	author       = {Yifang Zhu and Hinds, William Carson and Seongheon Kim and Si Shen and Constantinos Sioutas},
	year         = 2002,
	month        = sep,
	journal      = {Atmospheric Environment},
	volume       = 36,
	number       = 27,
	pages        = {4323--4335},
	doi          = {10.1016/S1352-2310(02)00354-0},
	issn         = {1352-2310}
}

\clearpage
\appendix

\section{Technical report}
\label{app:technical}

This appendix documents implementation details of the pipeline of Section~\ref{sec:methods} that are needed to reproduce it but are not needed to follow the method. Where a detail is specific to the case study, this is stated explicitly.

\subsection{Case-study fleet}
\label{app:tech-fleet}

The through-traffic fleet of the case study is inherited unchanged from the corridor demand model: four vehicle categories, namely motorcycles, small trucks, large trucks, and passenger cars, whose day-specific shares come from their detector statistics, with passenger cars further split into five behavioural subtypes reflecting different degrees of driving aggressiveness. In the route files generated for this study, the five passenger-car subtypes occur about equally often. This yields eight through-traffic vehicle types: a motorcycle, five passenger-car types differing only in car-following aggressiveness, a light delivery vehicle, and a truck. Randomised local traffic is generated at runtime as a single additional type (Appendix~\ref{app:tech-modules}). No HBEFA4 emission class is assigned explicitly to any of these types; \ac{SUMO} therefore resolves each type to a default class from its vehicle class (Section~\ref{sec:meth-emission}). The fleet is consequently uniform within each vehicle class and carries no representation of the age, fuel, or Euro-norm distribution of the actual vehicle population in the corridor.

\subsection{Optional traffic modules}
\label{app:tech-modules}

\paragraph{Randomised local traffic.} Randomised local traffic representing everyday resident car use can be injected at each simulation step. Spawn rates are derived from demographic and mobility attributes of the resident population and differ between weekdays and weekends. Spawn points are restricted to road edges with at least one lane below a configurable speed-limit threshold, which concentrates local traffic on lower-speed roads. From the set of eligible edges, a start and an end edge are drawn uniformly for every new vehicle, and its route is computed as a shortest path on the recent, smoothed travel times of the simulation, so that it responds to congestion. Local vehicles depart at speed zero with a random speed factor between 0.8 and 1.0.

In the case study, the population attributes are taken from SynPop~\citep{ARE2024SynPop}. For each village, the number of residents with a driving licence and access to a car was counted and split into employed residents, adjusted for part-time work, and residents who are neither employed nor in school but hold a licence and have a car. On weekdays, all employed residents are assumed to commute in the morning and in the evening, with half of them returning home for lunch; between 19 and 21\,h a share of all residents with a driving licence and a car drives, in the remaining daytime hours a share of the non-employed residents drives, and at night 0.1\,\% of residents are assumed to drive per hour. The weekend profile follows similar assumptions without the distinction between employed and non-employed residents. Eligible spawn edges have at least one lane with a speed limit below 17\,m\,s$^{-1}$ (about 61\,km\,h$^{-1}$) and either carry no vehicle-class restriction, allow \textit{passenger} vehicles, or allow \textit{delivery} vehicles while excluding pedestrians; this excludes edges restricted to a single other mode such as bicycles or buses.

Local traffic is spawned with a single vehicle type whose \ac{SUMO} vehicle class is set to \textit{delivery}, so that it can also use residential edges that \ac{OSM} restricts to delivery vehicles. Its emission class is not overridden and remains the \ac{SUMO} default, HBEFA4 passenger car, petrol, Euro~4 (\texttt{HBEFA4/PC\_petrol\_Euro-4}); all local vehicles therefore emit as petrol passenger cars regardless of their delivery routing class, uniformly across the fleet with no diesel, electric, or other Euro-standard share. These spawn rates and the uniform emission class are heuristic and were not calibrated against observed local traffic.

\paragraph{Exit-lane speed forcing.} Queues at the exit of a village depend on conditions outside the simulated domain. To carry this information in, the maximum speed of the exit-lane edges is overwritten at every simulation step with the speed read from a table, part of the traffic data we were given; before the first data point of the day, the first observed speed is used. Vehicles inside the domain queue behind the imposed speed, which itself does not respond to the simulated traffic.

\paragraph{Dynamic rerouting.} At configurable intervals, a specified share of vehicles recalculates the shortest path to its destination based on current simulated network conditions, reproducing the adaptive behaviour of drivers who switch routes in response to developing queues. In the case study, the interval is six minutes and the share is varied between 0 and 15\,\% (Section~\ref{sec:casestudy-scenarios}); no reference values for either quantity were found in the literature. The 5\,\% share used in Scenario~C was selected qualitatively on the basis of the visual plausibility of the resulting emission heat maps.

\subsection{Dispersion solver}
\label{app:tech-dispersion}

\paragraph{Grid.} Horizontal cells are approximately $10 \times 10$\,m, which is coarser than the $5 \times 5$\,m used by \textcite{zhao2024} for a pedestrian-scale district study but finer than the $50 \times 50$\,m used by \textcite{SONNENSCHEIN2025} for city-scale work, reflecting the intermediate spatial scale of a village domain. For the vertical direction we adopt the six-layer geometric grid of \textcite{zhao2024}, with face boundaries at $[0,\,0.3,\,1.2,\,4.8,\,19.2,\,76.8,\,200]$\,m spanning 0--200\,m. The inner boundaries follow a geometric progression with ratio~4, yielding layer thicknesses of 0.3, 0.9, 3.6, 14.4, 57.6, and 123.2\,m and concentrating resolution in the near-road zone where vertical concentration gradients are steepest. The lowest cell centre lies at 0.15\,m, the layer into which vehicles inject their emissions.

\paragraph{Vertical eddy diffusivity.} $K_z$ is evaluated at the vertical layer faces (0.3, 1.2, 4.8, 19.2, 76.8\,m, and the 200\,m top face, which enters the coupling to the top Dirichlet boundary), not at cell centres, so the finite-volume flux between layers $k$ and $k+1$ becomes
\begin{equation}
F_{k+1/2} = -K_z(z_{k+1/2})\,\frac{C_{k+1}-C_k}{z^c_{k+1}-z^c_k}.
\label{eq:kz-flux}
\end{equation}
Because $K_z$ depends only on $z$, every grid line in the $z$-sweep still shares a single tridiagonal matrix, so the multiple-right-hand-side banded solve is unchanged; only the matrix entries change from the constant $D_z$ to the face values of Eq.~\eqref{eq:kz}. $K_z(0)$ is undefined at the ground face; the flux there is unaffected, since the no-flux ground boundary condition sets it to zero directly rather than evaluating Eq.~\eqref{eq:kz-flux}. A boundary-layer shape factor such as $(1-z/z_i)^2$ was not used because it forces $K_z=0$ at the top Dirichlet face, which would decouple the top boundary from the interior.

\paragraph{Time integration.} Time is advanced with an implicit backward-Euler scheme using Lie splitting (sequential operator splitting, applied one spatial direction at a time). At each time-step the vehicle source $S^n$ is first deposited into the current concentration field, and the field is then updated in three sweeps, one per spatial direction:
\begin{align}
(I - \Delta t\,D_x\,L_x)\,C^{*} &= C^n + \Delta t\,S^n, \label{eq:adi-x}\\
(I - \Delta t\,D_y\,L_y)\,C^{**} &= C^{*}, \label{eq:adi-y}\\
(I - \Delta t\,L_z^{K})\,C^{n+1} &= C^{**}. \label{eq:adi-z}
\end{align}
Here $L_x$ and $L_y$ are standard second-order central-difference Laplacian operators on the uniform horizontal grid, and $L_z^{K}$ is the finite-volume operator $\partial_z(K_z\,\partial_z\,\cdot)$ on the geometric vertical grid, with $K_z$ evaluated at the layer faces (Eq.~\eqref{eq:kz-flux}). When the optional wind described below is enabled, $L_x$ and $L_y$ additionally carry a first-order upwind advection term; $L_z^{K}$ is unchanged, since there is no vertical advection. Because each operator in Eqs.~\eqref{eq:adi-x}--\eqref{eq:adi-z} couples a cell only to its two immediate neighbours along one direction, solving a full three-dimensional sweep reduces to solving many independent one-dimensional problems, one for each line of grid cells running in that direction. For the $x$-sweep, for example, each of the $N_y N_z$ lines of cells at fixed $(y,z)$ gives its own small system of equations, and because every cell in that system depends only on its immediate left and right neighbours, the system's matrix has non-zero entries only on the main diagonal and the two diagonals directly next to it (tridiagonal). Such systems can be solved in effort proportional to their size rather than to its cube, as a general matrix would require. Since the grid spacing and diffusion coefficient are the same on every line of a sweep, all $N_y N_z$ lines share the same matrix and differ only in their right-hand side, so the whole sweep reduces to one banded solve with $N_y N_z$ right-hand sides (respectively $N_x N_z$ and $N_x N_y$ for the $y$- and $z$-sweeps), computed with \texttt{scipy.linalg.solve\_banded}; transpositions and reshaping between sweeps are performed with NumPy. Backward Euler in each direction combined with Lie splitting makes the scheme first-order accurate in time and second-order in space.

\paragraph{Stability.} The implicit Lie-split scheme is unconditionally stable for all positive diffusion coefficients, so no \ac{CFL}-type restriction applies. The time-step is fixed at $\Delta t = 1$\,s to keep the per-vehicle emission injection temporally consistent with the one-second resolution of the SUMO microsimulation.

\paragraph{Boundary conditions.} The lateral boundaries, through ghost cells one cell width beyond the domain, and the top face are held at the ambient background concentration (Dirichlet), acting as an infinite reservoir that prevents artificial pollutant accumulation at the domain edges. In their systematic review of \ac{CFD} dispersion studies, \textcite{Pantusheva2022} find that most authors prescribe a velocity inlet and an outflow or pressure outlet for the flow; such flow conditions have no counterpart in a solver without a resolved flow, and we therefore hold the concentration itself at background on the lateral boundaries. The ground face applies a no-flux Neumann condition ($\partial C / \partial z \big|_{z=0} = 0$), implemented by omitting the downward flux term in the finite-volume balance of the lowest layer, reflecting the absence of a surface sink. After the source injection and again after the three sweeps, the concentration field is clipped from below at the ambient background value. This floor prevents the Dirichlet boundaries from acting as spurious sinks and keeps every cell at or above background, at the cost of a small non-conservative correction whenever a cell would otherwise fall below that level.

\paragraph{PM loss term.} The loss term $-\lambda\,(C - C_{\mathrm{bg}})$ for PM$_{2.5}$ is applied with backward Euler in the same step as the vertical sweep, which then becomes $(I - \Delta t\,L_z^{K} + \Delta t\,\lambda\,I)\,C^{n+1} = C^{**} + \Delta t\,\lambda\,C_{\mathrm{bg}}$; the matrix remains tridiagonal. $\lambda = 0$ for CO and NO$_2$. $\lambda$ acts uniformly across all six vertical layers, including those far above the road where little PM$_{2.5}$ reaches. A physical deposition process would instead act only at the surface, through a deposition velocity applied in the lowest layer; the uniform $\lambda$ used here is a calibration surrogate rather than a deposition model, consistent with the discussion of Section~\ref{sec:discussion}.

\paragraph{Optional uniform wind.} When enabled, a horizontal wind of prescribed speed and direction advects the concentration field. The wind is uniform over the domain and constant within each hour, and its direction is given as the compass bearing it blows from. It is disabled in all calibration runs and in all results except Section~\ref{sec:windsens}. Advection is discretised with a first-order upwind, implicit (backward Euler) finite-volume scheme. For a uniform, divergence-free wind on the uniform horizontal grid the upwind difference coincides with the conservative flux form, so the scheme conserves mass away from the domain boundaries. Advection is not a separate operator: it is added into the horizontal diffusion sweeps of Eqs.~\eqref{eq:adi-x}--\eqref{eq:adi-y}, so that each row of the $x$- and $y$-sweeps solves diffusion and advection together in one tridiagonal system; there is no vertical advection. At the lateral Dirichlet boundaries, the upstream ghost cell holds the ambient background, so inflow enters at background concentration; the downstream boundary carries no advective term, only the diffusive coupling to the background ghost cell, so the plume leaves the domain freely by advection while a small diffusive pull toward background remains. The scheme is unconditionally stable at any Courant number, so $\Delta t$ stays at 1\,s regardless of wind speed; accuracy, not stability, is the limit.

\paragraph{Implementation.} Grids are initialised at the ambient background concentration for each pollutant, so that the simulation starts from a realistic state; because runs begin at midnight, when traffic is near-zero, this initialisation is self-consistent. All runs were executed on CPU cores only, and the solver was tested for mass conservation, monotone decay, and agreement with the analytical Gaussian solution. All runs use a fixed random seed for the simulator and for the random number generators of the local-traffic and rerouting modules, so a run is reproducible; each scenario was simulated once. Spawning and rerouting draw from separate random streams, so the local-traffic demand is identical across the rerouting shares of the sweep in Section~\ref{sec:casestudy-scenarios}.

\subsection{AQI evaluation}
\label{app:tech-aqi}

The sub-index for each pollutant is evaluated one layer above the ground, at 0.3 to 1.2\,m, closer to breathing height than the layer into which vehicles inject their emissions, and converted to the units of the \ac{EPA} breakpoint tables (ppm for CO, ppb for NO$_2$, and $\mu$g\,m$^{-3}$ for PM$_{2.5}$) using a molar volume of 24.45\,L\,mol$^{-1}$ (25\textdegree C, 1\,atm). The simulation saves hourly grid snapshots rather than a continuous record, so the 1-hour NO$_2$ sub-index reduces to the instantaneous concentration at each saved hour, and the 8-hour and 24-hour windows are means of these hourly snapshots rather than continuously integrated averages. For NO$_2$ every saved hour is its own window, for CO the 8-hour windows overlap, and for PM$_{2.5}$ a simulated day contains exactly one 24-hour window. The sub-index of a resident for a given pollutant and day is the maximum value found across all windows of that day, so that the worst case is reported rather than one fixed window.

\section{Dispersion parameters}
\label{app:params}
 
The simulation covers the roads inside the villages, not the A2, so the concentrations it produces should lie between the cleanest and the most polluted air measured in the valley. The background concentration of each pollutant is therefore set to the lowest value measured in the Uri Valley between Altdorf and Göschenen, and the near-road target to the highest. Where the valley has no measurement, values from external sources are used, as indicated per cell. The horizontal diffusion coefficients $D_x = D_y$ and, for PM$_{2.5}$, a first-order loss rate $\lambda$ were chosen by a scripted search so that simulated concentrations are of the same order as the near-road targets and decay over roughly the distances in Table~\ref{tab:params}; the vertical diffusivity follows the Monin--Obukhov parameterisation of Section~\ref{sec:meth-dispersion} instead of a hand-tuned constant. Calibration replayed recorded traffic through the dispersion solver without wind, over a period of heavy congestion on the high-traffic day (20 July 2024), and evaluated concentrations in the 0.3--1.2\,m exposure layer.
 
\begin{table}[H]
\centering
\caption{\textbf{Background concentration, near-road target and decay distance} for each pollutant, with the source given in each cell.}
\label{tab:params}
\renewcommand{\arraystretch}{1.3}
\setlength{\tabcolsep}{4pt}
\scriptsize
\begin{tabularx}{\textwidth}{@{}>{\RaggedRight\arraybackslash}p{1.3cm} Y{1} Y{1} Y{1}@{}}
\toprule
& \textbf{Background (used directly)} & \textbf{Near-road (target)} & \textbf{Decay distance (target)} \\
\midrule
NO$_2$ &
\textbf{5.1\,$\mu$g\,m$^{-3}$}: lowest value in the valley, rural background site Altdorf Kapuzinerkloster, 270\,m from the nearest road~\citep{AfUUri2025} &
\textbf{24.4\,$\mu$g\,m$^{-3}$}: highest value in the valley, Altdorf Gross Ei, 2.5\,m from the A2~\citep{AfUUri2025} &
\textbf{${\sim}300$\,m}: along the A2, 24.4\,$\mu$g\,m$^{-3}$ at 2.5\,m, 16.6 at 5\,m, 13.2 at 100\,m, 7.6 at 270\,m and 7.2 at 700\,m~\citep{AfUUri2025} \\
\addlinespace[3pt]
PM$_{2.5}$ &
\textbf{5.5\,$\mu$g\,m$^{-3}$}: only PM$_{2.5}$ value in the valley, station A2 Uri, 5\,m from the A2, so lowest and highest coincide~\citep{AfUUri2025} &
\textbf{${\sim}7.5$\,$\mu$g\,m$^{-3}$}: no valley data; background plus the regression-modelled near-road increment across US near-road sites, about 2\,$\mu$g\,m$^{-3}$ at 5\,m~\citep{Mukherjee2020} &
\textbf{${\sim}30$\,m}: no valley data; the same increment falls to about 0.5\,$\mu$g\,m$^{-3}$ at 30\,m~\citep{Mukherjee2020} \\
\addlinespace[3pt]
CO &
\textbf{0.17\,mg\,m$^{-3}$}: no valley data; lowest background mixing ratio at the suburban \ac{NABEL} site Dübendorf, 150\,ppb, measured 2002--2005~\citep{Steinbacher2007} &
\textbf{0.175\,mg\,m$^{-3}$}: no valley data; value at the road ($x=0$) of the pooled fit across three Texas roadways under perpendicular wind, $y = 61.5 + 91.3\exp(-0.0151x)$ ($R^2 = 0.73$), i.e.\ 152.8\,ppb~\citep{Clements2009} &
\textbf{${\sim}300$\,m}: no valley data; CO returned to background within a few hundred metres of three Texas roadways~\citep{Clements2009} \\
\bottomrule
\end{tabularx}
\end{table}

\paragraph{Horizontal diffusion and PM loss.} The coefficients were found by searching over $D_x=D_y$, and for PM$_{2.5}$ additionally $\lambda$, with the vertical diffusivity held fixed, until a score comparing the near-road concentration and its decay with the targets above was minimised; the achieved values are not reported and no agreement with the targets is claimed. CO and NO$_2$ share $D_x=D_y=640\,\mathrm{m^2\,s^{-1}}$ with $\lambda=0$; PM$_{2.5}$ uses $D_x=D_y=25\,\mathrm{m^2\,s^{-1}}$ with $\lambda=0.02\,\mathrm{s^{-1}}$. The gas value exceeds the vertical diffusivity of Eq.~\eqref{eq:kz} near the ground (about $2.4\,\mathrm{m^2\,s^{-1}}$ at 5\,m) by more than two orders of magnitude. It is therefore an effective coefficient that stands in for the horizontal transport by wind that the main model omits, not a physical diffusivity. The horizontal coefficients are tuning constants chosen to give plausible concentration levels and decay lengths, so the 25-fold difference between the gases and PM$_{2.5}$ has no physical interpretation. Likewise, $\lambda$ corresponds to a lifetime of about 50\,s for the traffic-attributable excess and is a calibration surrogate, not a measured deposition rate. As demonstrating the pipeline end-to-end is the contribution of this paper rather than a calibrated air-quality forecast, these parameters are reported as the fixed values used throughout the case study rather than as a formal fit.

\paragraph{Vertical diffusivity.} The Monin--Obukhov inputs of Eq.~\eqref{eq:kz} are matched to \textcite{zhao2024}: $L = -200$\,m, $z_0 = 0.1$\,m, and $u_* \approx 0.91\,\mathrm{m\,s^{-1}}$.\footnote{Our own estimate, computed from \textcite{zhao2024}'s stated 10\,m\,s$^{-1}$ wind via the stability-corrected logarithmic wind profile, assuming the wind is measured at $z_{\mathrm{ref}} = 5$\,m.}

\paragraph{Ambient floor.} Through the \ac{EPA} breakpoints of Eq.~\eqref{eq:aqi}, the background concentrations of Table~\ref{tab:params} map to sub-indices of 2.56 for NO$_2$, 30.56 for PM$_{2.5}$ and 1.69 for CO. Their sum, $I_{\mathrm{bg}} = 34.81$ points, is the ambient floor of Section~\ref{sec:meth-exposure} that is subtracted to obtain the excess index $\Delta I$.

\section{Example heat-map output}
\label{app:heatmaps}

The figures below show the concentration field at 6\,PM on 20~July (Scenario~C) in the 0.3--1.2\,m layer, with one figure per pollutant and one panel per village, ordered from north to south as in Section~\ref{sec:casestudy-area}. The lower end of each colour scale is the ambient background concentration of that pollutant (Appendix~\ref{app:params}), so any colour above the lightest shade is a contribution from simulated traffic. The upper end is set near the highest concentration reached at any residential building across the six villages, so that dwelling values remain resolved. Roads can exceed the upper end and are then also shown in the top colour. Green polygons mark residential buildings. Spatial extents differ between panels because each domain covers only the residential area of its village (Figure~\ref{fig:all_villages_squares}). These are single-hour snapshots and not the time-averaged quantities behind the \ac{AQI} of Section~\ref{sec:res-villages}, so they illustrate spatial structure rather than reproduce the exposure ranking.

Two features are visible across all three figures. First, the plumes differ strongly in width: NO$_2$ and CO spread well beyond the road, whereas PM$_{2.5}$ stays confined to a narrow band along it, a direct consequence of the much smaller horizontal diffusion coefficient and the loss term chosen for PM$_{2.5}$ to reproduce its much shorter decay distance (Appendix~\ref{app:params}). Second, the road-level maximum and the residential exposure are decoupled. Göschenen has the highest road-level CO of the six domains, yet its dwellings lie west of the road and outside the plume, and their mean concentration is barely above background. Wassen shows the opposite: its road maximum is lower, but its buildings lie inside the plume around the motorway exit, and their mean is the highest of the six villages.

\begin{figure}[H]
\centering
\includegraphics[width=\linewidth]{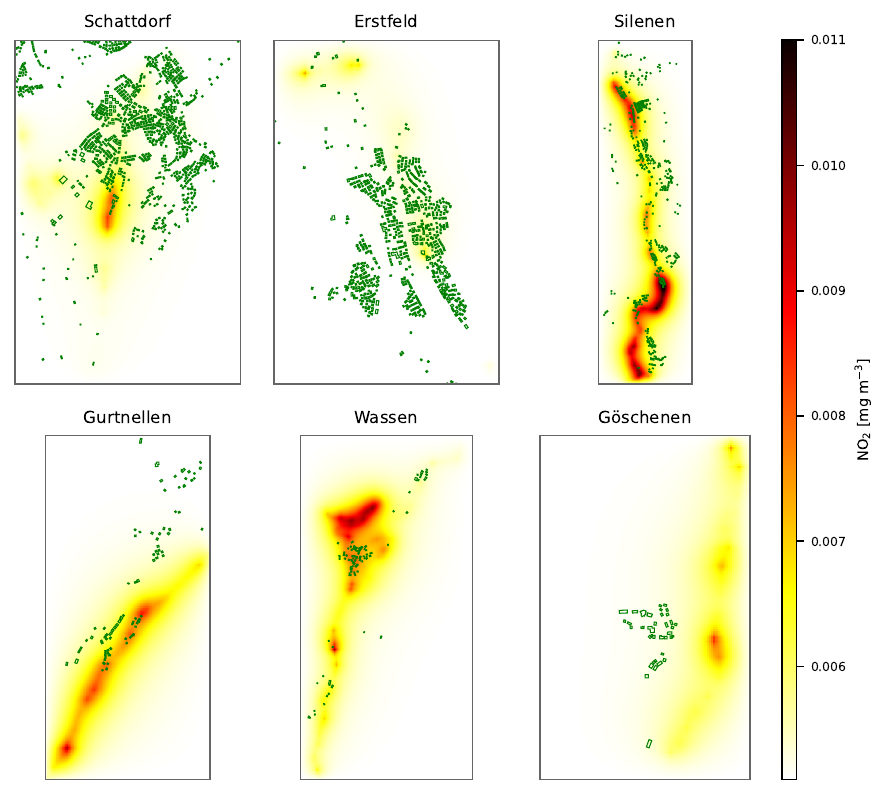}
\caption{\textbf{NO$_2$ concentration} at 6\,PM on 20~July (Scenario~C), shared colour scale from the ambient background upward. The highest residential exposure occurs where a road runs through the built-up area: around the motorway exit in Wassen and along the cantonal road in Silenen and Gurtnellen. Schattdorf shows a single moderate hotspot on the main road, Erstfeld is barely above background at this hour, and in Göschenen the elevated band follows the road east of the dwellings.}
\label{fig:app_heatmaps_no2}
\end{figure}

\begin{figure}[H]
\centering
\includegraphics[width=\linewidth]{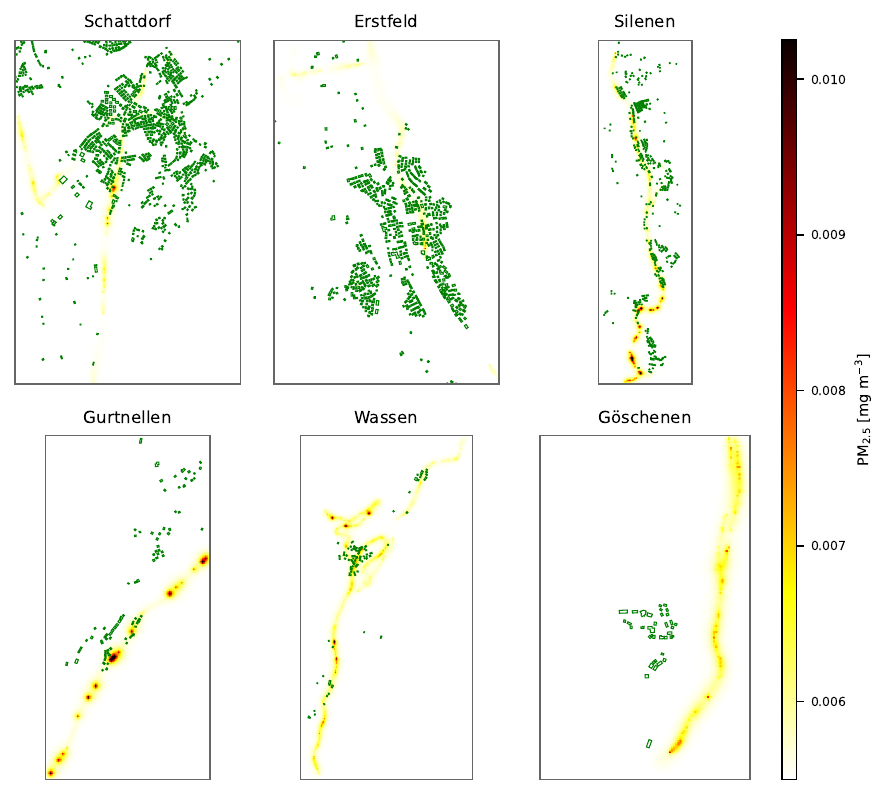}
\caption{\textbf{PM$_{2.5}$ concentration} at 6\,PM on 20~July (Scenario~C), shared colour scale from the ambient background upward. The plume is far narrower than for NO$_2$ and traces the road almost exactly, with isolated peaks along it. Dwellings are reached mainly where they sit directly beside the road, as in Silenen, Gurtnellen and Wassen.}
\label{fig:app_heatmaps_pm25}
\end{figure}

\begin{figure}[H]
\centering
\includegraphics[width=\linewidth]{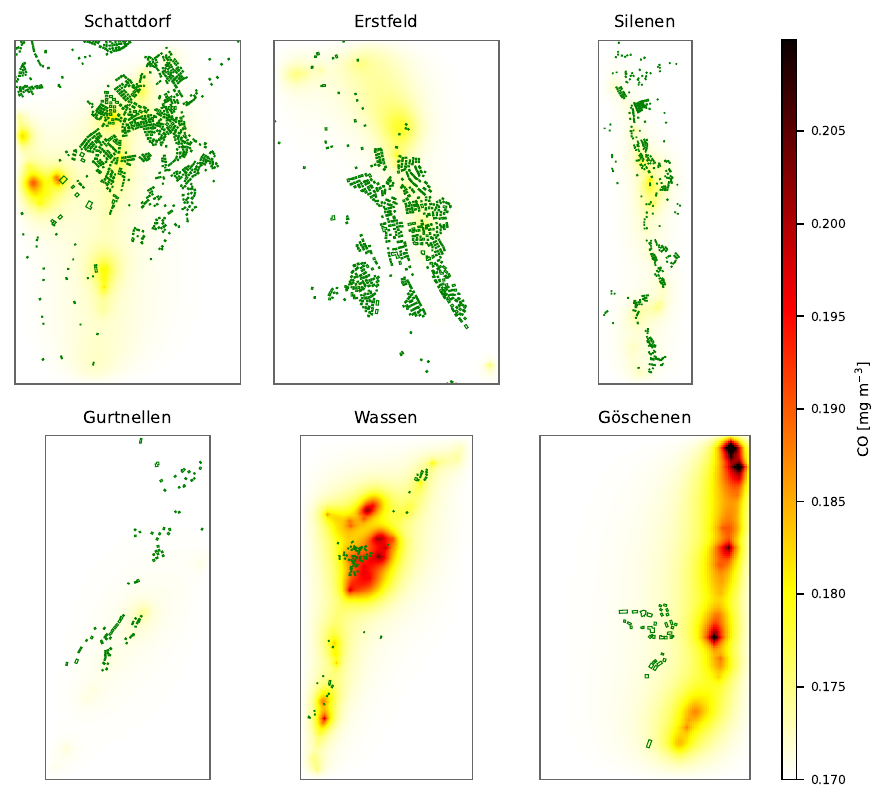}
\caption{\textbf{CO concentration} at 6\,PM on 20~July (Scenario~C), shared colour scale from the ambient background upward. The elevated areas are broader than for PM$_{2.5}$, but at this hour CO reaches dwellings mainly in Wassen, where a hotspot covers the village centre. The highest road-level values occur in Göschenen, whose dwellings lie west of the road; Erstfeld and Gurtnellen stay close to background.}
\label{fig:app_heatmaps_co}
\end{figure}

\section{Per-pollutant bar plots}
\label{app:barplots}

Figure~\ref{fig:aqi_summary} reports the summed sub-indices. The figures below break that sum into its three components, with the same layout throughout: unweighted sum of the sub-index over all residents (left), mean sub-index per resident (centre), and mean excess sub-index per resident (right), for the high-traffic day of 20~July and the low-traffic reference day of 20~November under Scenario~C. Each pollutant has its own background floor, namely 2.56 index points for NO$_2$, 30.56 for PM$_{2.5}$ and 1.69 for CO (Appendix~\ref{app:params}), and the right-hand panel of each figure is the left-over after subtracting it.

Read together, the three figures show that the decomposition argued for in Section~\ref{sec:res-villages} is needed for every pollutant. Even for NO$_2$, the excess approaches its background floor only in Wassen on the busy day; elsewhere it stays well below it, so the raw means differ little between villages except for Wassen. For PM$_{2.5}$ and CO the floor dominates even more: the raw means are nearly flat across all six villages and both dates, while the excess differs by a large factor between the busy and the quiet day in most villages. Without the excess panels, a reader of these plots would conclude that the busy and the quiet day are much alike everywhere except in Wassen.

The second common feature is that Wassen has the largest mean excess of the six villages on the busy day for every one of the three pollutants, not only for the NO$_2$ that dominates the combined index. Because the three pollutants have different sources within the same vehicle fleet and different diffusion behaviour (Appendix~\ref{app:heatmaps}), this points to settlement geometry rather than to any pollutant-specific mechanism: the motorway exit discharges into the built-up area, so residents sit inside the plume whatever is in it. On the quiet day, Schattdorf has the largest excess for NO$_2$ and CO, consistent with its weekday local traffic, while Wassen remains highest for PM$_{2.5}$. Göschenen shows the reverse of the usual pattern for PM$_{2.5}$: its mean excess is markedly higher on the quiet day than on the busy day, because on the busy day the domain carries mostly through vehicles whose narrow PM$_{2.5}$ plume stays on the road east of the dwellings, whereas on the quiet day the weekday local traffic drives next to the dwellings.

\begin{figure}[H]
\centering
\includegraphics[width=\linewidth]{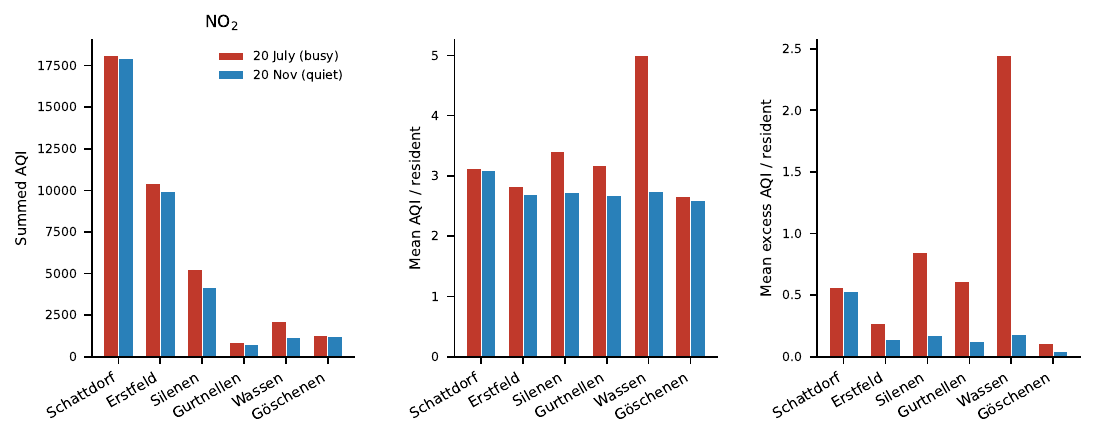}
\caption{\textbf{Per-village NO$_2$ AQI exposure} under Scenario~C. Layout as in Figure~\ref{fig:aqi_summary}: unweighted sum (left), mean per resident (centre), mean excess per resident (right). NO$_2$ has the largest excess of the three pollutants. Wassen stands apart on the busy day, while Schattdorf's excess is nearly the same on both days.}
\label{fig:app_barplots_nox}
\end{figure}

\begin{figure}[H]
\centering
\includegraphics[width=\linewidth]{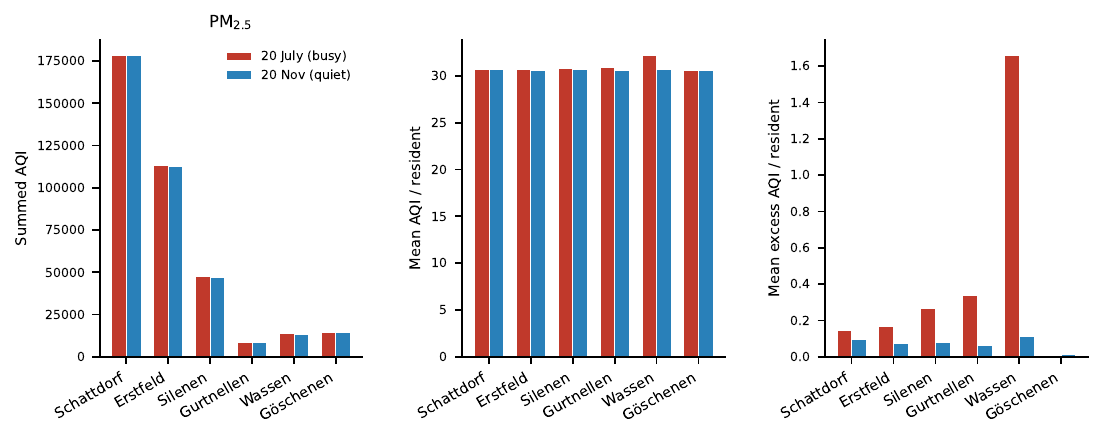}
\caption{\textbf{Per-village PM$_{2.5}$ AQI exposure} under Scenario~C, layout as above. PM$_{2.5}$ has the largest sub-index of the three in absolute terms, which is why the sub-indices are summed rather than reduced to their maximum (Section~\ref{sec:meth-exposure}), but almost all of it is the background floor. The centre panel is correspondingly flat, with every village close to that floor on both dates, while the right panel shows Wassen well above the rest on the busy day.}
\label{fig:app_barplots_pmx}
\end{figure}

\begin{figure}[H]
\centering
\includegraphics[width=\linewidth]{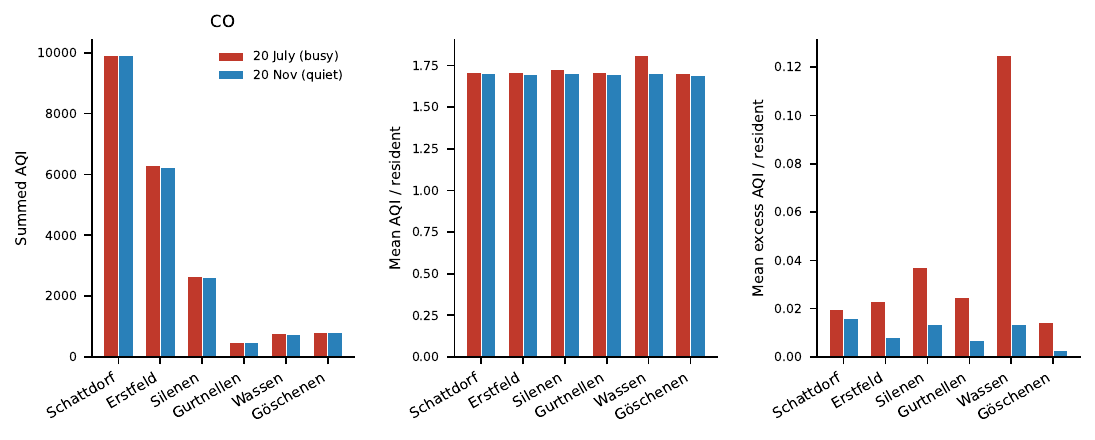}
\caption{\textbf{Per-village CO AQI exposure} under Scenario~C, layout as above. CO contributes the smallest excess of the three pollutants in all villages except Göschenen on the busy day. Its raw mean sits just above its floor in every village except Wassen on the busy day, and the summed panel is dominated by population size rather than by traffic.}
\label{fig:app_barplots_co}
\end{figure}

\end{document}